\documentclass[twocolumn,twocolappendix]{aastex63}
\usepackage{amsmath}
\hypersetup{linkcolor=blue,citecolor=blue,urlcolor=blue}
\usepackage{xcolor}
\usepackage[normalem]{ulem}
\definecolor{dgreen}{RGB}{26,148,49}

\DeclareRobustCommand{\ion}[2]{%
\relax\ifmmode
\ifx\testbx\f@series
{\mathbf{#1\,\mathsc{#2}}}\else
{\mathrm{#1\,\mathsc{#2}}}\fi
\else\textup{#1\,{\mdseries\textsc{#2}}}%
\fi}

\shorttitle{Phase-Space Spectral Line De-confusion in Intensity Mapping}
\shortauthors{Cheng, Chang and Bock}

\begin{document}

\title{Phase-Space Spectral Line De-confusion in Intensity Mapping}

\author[0000-0002-5437-0504]{Yun-Ting Cheng}
\address{California Institute of Technology, 1200 E. California Boulevard, Pasadena, CA 91125, USA}
\email{ycheng3@caltech.edu}

\author[0000-0001-5929-4187]{Tzu-Ching Chang}
\address{California Institute of Technology, 1200 E. California Boulevard, Pasadena, CA 91125, USA}
\address{Jet Propulsion Laboratory, California Institute of Technology, 4800 Oak Grove Drive, Pasadena, CA 91109, USA}
\address{Institute of Astronomy and Astrophysics, Academia Sinica, 1 Roosevelt Road, Section 4, Taipei, 10617, Taiwan}

\author{James J. Bock}
\address{California Institute of Technology, 1200 E. California Boulevard, Pasadena, CA 91125, USA}
\address{Jet Propulsion Laboratory, California Institute of Technology, 4800 Oak Grove Drive, Pasadena, CA 91109, USA}

\begin{abstract}
Line intensity mapping (LIM) is a promising tool to efficiently probe the three-dimensional large-scale structure by mapping the aggregate emission of a spectral line from all sources that trace the matter density field. Spectral lines from different redshifts can fall in the same observed frequency and be confused, however, which is a major challenge in LIM. In this work, we develop a line de-confusion technique in map space capable of reconstructing the three-dimensional spatial distribution of line-emitting sources. If multiple spectral lines of a source population are observable in multiple frequencies, using the sparse approximation, our technique iteratively extracts sources along a given line of sight by fitting the LIM data to a set of spectral templates.  We demonstrate that the technique successfully extracts sources with emission lines present at a few $\sigma$ above the noise level, taking into account uncertainties in the source modeling and presence of continuum foreground contamination and noise fluctuations. As an example, we consider a TIME/CONCERTO-like survey targeting [\ion{C}{ii}] at the epoch of reionization, and reliably reconstruct the 3D spatial distribution of the CO interlopers and their luminosity functions at $0.5\lesssim z\lesssim 1.5$.  We also demonstrate a successful de-confusion for the SPHEREx mission in the near-infrared wavelengths.We discuss a formalism in which the reconstructed maps can be further cross-correlated with a (galaxy) tracer population to estimate the total interloper power. This technique is a general framework to extract the phase-space distribution of low-redshift interlopers, without the need of external information, for any line de-confusion problem.
\footnote{\textcircled{c} 2020. All rights reserved.}
\end{abstract}

\section{Introduction}
Line intensity mapping (LIM) has emerged as a promising tool to study the three-dimensional large-scale structures by mapping a particular spectral line emission, and infers the line-of-sight distance of the emission sources from the frequency-redshift relation. LIM measures the aggregate emission of all sources to constrain the bulk properties of the galaxies; whereas in traditional galaxy surveys, only the brighter sources can be individually detected. This relatively low spatial resolution and point-source sensitivity requirement in LIM enables the use of small apertures to efficiently scan a large survey volume out to high redshifts. 

Several spectral lines have been proposed for LIM survey. The 21 cm hyperfine emission from neutral hydrogen  \citep{1990MNRAS.247..510S,1997ApJ...475..429M,2008PhRvL.100i1303C,2008MNRAS.383..606W}, the CO rotational lines \citep{2008A&A...489..489R,2010JCAP...11..016V,2011ApJ...730L..30C,2011ApJ...741...70L, 2011ApJ...728L..46G,2014MNRAS.443.3506B, 2013ApJ...768...15P, 2015JCAP...11..028M,2015ApJ...814..140K,2016MNRAS.457L.127B,2016ApJ...817..169L,2016ApJ...830...34K,2017MNRAS.464.1948F,2017MNRAS.468..741B,2019ApJ...872..186C}, the [\ion{C}{ii}] 157.7 $\mu$m fine-structure line \citep{2012ApJ...745...49G,2014ApJ...793..116U,2015ApJ...806..209S,2015MNRAS.450.3829Y,2017MNRAS.464.1948F}, and the Lyman-$\alpha$ emission line \citep{2013ApJ...763..132S,2014ApJ...785...72G,2014ApJ...786..111P, 2016MNRAS.455..725C,2016MNRAS.457.3541C,2017MNRAS.464.1948F,2018MNRAS.481.1320C} are amongst the most studied lines in the LIM regime.

One of the main challenges in LIM is the astrophysical foreground contaminations, including the continuum emission and line interlopers. Although the continuum foregrounds are usually a few orders of magnitude brighter than the lines (a situation more severe for 21cm than for other lines), their smooth spectral feature can be used to distinguish from the line signals. This has been extensively studied in the context of 21 cm LIM \citep[e.g.,][]{2006PhR...433..181F,2006ApJ...648..767M,2009ApJ...695..183B,2012MNRAS.419.3491L,2012ApJ...756..165P,2012MNRAS.423.2518C,2015ApJ...815...51S}. The line interlopers, which originate from sources residing in different redshifts emitting spectral lines in the same observed frequency channel, is another pressing issue for LIM experiments. The two most studied line de-confusion techniques, source masking and cross-correlation, typically rely on external data sets that trace the same cosmic volume: the masking technique makes use of a galaxy survey catalog to identify and remove bright interloper sources \citep{2015MNRAS.452.3408B,2015MNRAS.450.3829Y,2015ApJ...806..209S,2018ApJ...856..107S}, whereas cross-correlation of an LIM survey with an external (or internal)  data set can help extract signals of interest \citep{2009ApJ...690..252L,2010JCAP...11..016V,2012ApJ...745...49G, 2014ApJ...785...72G,2015aska.confE...4C,2015ApJ...806..209S,2010Natur.466..463C,2013ApJ...763L..20M, 2016MNRAS.457.3541C,2019ApJ...872..186C}.  In addition to these two methods, \citet{2016ApJ...825..143L} and \citet{2016ApJ...832..165C} use the anisotropy of the interloper power spectrum arising from projection to the target line redshift to separate the lines. \citet{2020ApJ...894..152G} distinguish the lines from the same projection effect but using the multipole power spectrum. \citet{2014arXiv1403.3727D} propose to use angular fluctuations of the light to reconstruct the 3D source luminosity density.

Most of the existing line de-confusion methods (e.g. cross-correlation and power spectrum anisotropy) only extract the two-point statistics  (power spectrum or correlation function) but lose the phase information of individual line maps, which are valuable for cosmological parameter constraints and systematics control in the data. With individual line maps, one can extract information beyond two-point statistics in the non-Gaussian intensity maps, especially ones from the epoch of reionization (EoR). For example, \citet{2017MNRAS.467.2996B} and \citet{2019ApJ...871...75I} show that the one-point statistics of the intensity field can help constrain the luminosity function model. In addition, individual line maps can be used directly as density tracers for various cross-correlation, multi-tracer analysis, de-lensing of the cosmic microwave background (CMB), and perform consistency tests on different spatial regions with different foreground properties.

In this work, we develop a technique to extract individual line intensity maps from an LIM data set with blended interlopers. Using the fact that when multiple spectral lines emitted by a source are observable in an LIM survey, the redshift of the source can be pinned down by fitting to a set of spectral templates that are unique at each redshift. Without any external tracers or spectroscopic follow-up observations, individual line maps can be directly derived. For demonstration, we apply our technique to simulated data of an LIM survey targeting the EoR [\ion{C}{ii}] line with multiple low-redshift CO interlopers. In this case, the intensity field of the low-$z$ CO lines ($0.5\lesssim z\lesssim 1.5$) can be reconstructed since they can be detected in multiple spectral channels.

\citet{2015ApJ...806..234K} first explored the map-space line de-confusion using the multi-line wavelength information in the context of a pencil-beam spectroscopic survey. In this work, we explore the technique in the LIM regime that has a much lower sensitivity and the spectral resolution. In this regime, our template-fitting-based technique can obtain the signal-to-noise ratio (S/N) of the desired signals by 
using the data from multiple frequency channels.

The recent work by \citet{2020MNRAS.496L..54M} demonstrated the feasibility of LIM phase-space de-confusion with deep learning. They show that, in the absence of noise or foreground components, their algorithm can reconstruct the individual line maps that are mixed in the LIM data set. Their training data generation relies on the assumption of the signal clustering and the line luminosity model, whereas in this work, we develop a line de-confusion technique that only makes use of the spectral feature of the lines, which is more robust against the model uncertainty and the noise.

This paper is organized as follows. First we introduce the model and the survey parameters we used to generate the mock LIM survey data in Sec.~\ref{S: model}. Then, Sec.~\ref{S:methods} describes our line de-blending technique. Sec.~\ref{S:results} presents the results on the fiducial setup. In Sec.~\ref{S:discussion}, we present the performance of the technique with more practical considerations, and discuss its applications and extensions. The conclusion are given in Sec.~\ref{S:conclusion}. Throughout this paper, we consider a flat $\Lambda$CDM cosmology with $n_s=0.97$, $\sigma_8=0.82$, $\Omega_m=0.26$, $\Omega_b=0.049$, $\Omega_\Lambda=0.69$, and $h=0.68$, consistent with the measurement from Planck \citep{2016A&A...594A..13P}.

\section{Mock Light Cone Construction}\label{S: model}
For each spatial pixel in an LIM survey, there are a set of spectral channel measurements. Hereafter, the term ``light cone''  refers to the collection of the spectral measurements in a single pixel. The line de-confusion method introduced in this work is performed on a pixel-by-pixel basis, which only utilizes the spectral information in an individual line of sight (light cone), without taking into account the spatial clustering information, which we leave for future work.

We test our de-confusion technique on simulated light cones. Since the clustering information is not relevant to our technique, we generate light cones that are based solely on the spectral line luminosity function models and not on the clustering properties. Thus, all light cones are independent from one another, and we also ignore the line-of-sight clustering in this work. This allows for both the light cone construction and de-confusion by parallelization to speed up without affecting the quantification of performance.

As a demonstration of the technique, we assume an LIM experiment targeting the redshifted [\ion{C}{ii}] fine-structure emission from the EoR: the LIM data set contains multiple low-$z$ CO rotational transitions from low redshifts as interlopers. We note that the technique can be readily applied to any line-confusion problem at other wavelengths.

\subsection{Line Signal Models}\label{S:model_sig}
We model the line emissions of the redshifted [\ion{C}{ii}] emission and five low-redshift CO $J$-transitions: \{CO(2--1), CO(3--2), CO(4--3), CO(5--4), CO(6-5)\}. In reality, in the sub-mm spectral range of interest (generally in the $\sim$ 200-300 GHz range), there are higher CO $J$ lines (that are fainter), Galactic and extragalactic dust continuum emissions (the cosmic infrared background), CMB radiation, and atmospheric emissions that can all contribute to the measurements. Since none of them will produce strong spectral features that impact the performance of our technique, we will not include them in the light cones. Instead, in Sec.~\ref{S:bg}, we will demonstrate that the continuum foreground mitigation in the data analysis process has negligible a impact on our technique performance.

We use the [\ion{C}{ii}] and CO luminosity function models provided by \citet{2016MNRAS.461...93P}, in which a semi-analytic model including the effect of radiative transfer was used to estimate the CO and [\ion{C}{ii}] luminosity functions as constrained by current observations. Here we adopt their fitted Schechter luminosity functions to construct our light cones.

\subsection{Survey Parameters}\label{S:survey_params}
We consider a mock experiment that has similar survey parameters as the two ongoing EoR [\ion{C}{ii}] LIM experiments, TIME \citep{2014SPIE.9153E..1WC} and CONCERTO \citep{2018A&A...609A.130L}. 

The mock survey covers 200--305 GHz with 70 evenly spaced spectral channels ($\delta\nu = 1.5$ GHz), and the $\Omega_{\rm pix} = 0.43^2$ arcmin$^2$ pixel size. We assume the instrument noise is white and has a Gaussian distribution, with four different per-pixel noise levels of standard deviation $\sigma_n=$ \{$10^3$, $5\times10^3$, $10^4$, $5\times10^4$\} Jy sr$^{-1}$. These values are comparable to the range of expected instrument noise in TIME and CONCERTO. 

\subsection{Light Cone Generation}\label{S:model_lightcone}
\begin{figure*}[ht!]
\begin{center}
\includegraphics[width=\linewidth]{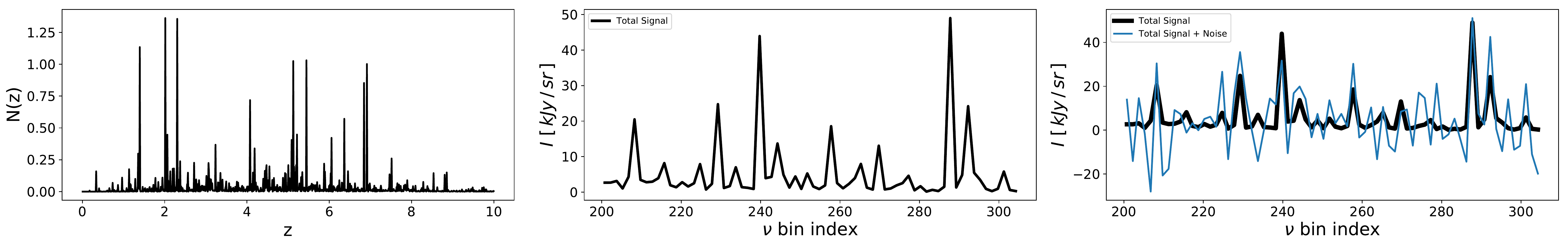}
\caption{\label{F:gen_lc}The steps of constructing a light cone. Left: $N(z)$, the effective number of $\ell_*$ sources in each redshift bin in this light cone. Middle: the signals from all six spectral lines and the sources in the 70 spectral channels. Right: the mock observed light cone (blue) consists of the signals (middle panel) and the $\sigma_n = 10$ kJy Gaussian noise. The black line is the signal component (same as the middle panel) for reference.}
\end{center}
\end{figure*}

Based on the assumed luminosity function models and survey parameters, we populate the light cones with sources drawn from random realizations of the Schechter function model.  We first define a fiducial Schechter function \footnote{For the fiducial Schechter function, we choose the $\phi_*$ and $\alpha$ values of CO(1--0) for $z\leq5$ and [\ion{C}{ii}] for $z>5$
in \cite{2016MNRAS.461...93P}. We interpolate or extrapolate the Schechter function parameters to the desired redshift from the values of $z=[0,1,2,3,4,6]$ given by \cite{2016MNRAS.461...93P}.
} for all of the lines as a function of redshift, 
\begin{equation}\label{E:lfunc}
\Phi(\ell)=\phi_\ast\, (\mathcal{L}/\mathcal{L}_*)^\alpha\, e^{-(\mathcal{L}/\mathcal{L}_*)}.
\end{equation}
We discretize the luminosity and redshift into bins of $\Delta (\mathcal{L}/\mathcal{L}_*)$ in 100 luminosity bins in log-space from $\mathcal{L}/\mathcal{L}_*=10^{-3}$ to $10$, and 2000 redshift bins of $\Delta z=5\times10^{-4}$ in linear space from $z=0$ to $z=10$.  
The expectation value of the source counts $N$ within each $\Delta \mathcal{L}$ and $\Delta z$ bin is given by
\begin{equation}
\langle N\rangle = \Phi(\ell) [\Delta (\mathcal{L}/\mathcal{L}_*)] [\Omega_{\rm pix} D_A^2(z) \frac{d\chi}{dz}\Delta z],
\end{equation}
where the last square bracket is the co-moving volume of the voxel defined by spatial pixel $\Omega_{\rm pix}$ and redshift bin $\Delta z$, $D_A$ is the co-moving angular diameter distance, and $\chi$ is the co-moving distance. 

For each $(\mathcal{L}, z)$ bin, we assign its source counts to a Poisson random number with expectation value $\langle N\rangle$, and then integrate along $\mathcal{L}$ to get the total luminosity in each redshift bin $\mathcal{L}_{\rm tot}(z)$. We define $N(z) \equiv \mathcal{L}_{\rm tot} / \mathcal{L}_*$ as the ``effective number counts'' per redshift bin. The left panel of Fig.~\ref{F:gen_lc}  shows the $N(z)$ of one example light cone.

Next, we assign the line luminosity signals in each redshift bin to $N(z)\,\ell_*^{\rm line}$, where $\ell_*^{\rm line}$ is the $\ell_*$ value of the line in the \citet{2016MNRAS.461...93P} model. We  then project the line signals to their corresponding spectral channels to make the light cones. The middle panel of Fig.~\ref{F:gen_lc} shows an example light cone spectrum from the $N(z)$ shown in the left panel.

Finally, we add a Gaussian random fluctuation with a an rms $\sigma_n$ value to each channel to account for the instrumental noise. The right panel of Fig.~\ref{F:gen_lc} shows the same light cone with a $\sigma_n = 10$ kJy Gaussian noise.

This light cone construction procedure assumes that all of the spectral lines have the same Schechter function parameters ($\phi_*$ and $\alpha$) and thus the same luminosity function shape, and the CO spectral line energy distribution (SLED) is also fixed. That is, by construction, all of the sources have the same line luminosity ratio sets by the relative value of $\ell_*^{\rm line}$. In the main parts of this work, we use this fixed SLED model, and we test the impact of adding the SLED variation on the performance of our technique in Sec.~\ref{S:model_uncertainty}.

We further point out that, even though the luminosity function has been sampled to the faint end ($10^{-3}\ell_*$), the light cone signals are still dominated by a few bright peaks. This indicates that the emission field can be well described by the few bright sources in the data. In other words, to extract the line emission field from a single line, one only needs to determine the redshift and luminosity of those bright sources. This is the main concept of our de-confusion technique, detailed in Sec.~\ref{S:methods}.

\section{Methods}\label{S:methods}
\subsection{Formalism}\label{S:methods_matrix}

The intensity of a light cone in frequency channel $\nu_i$ can be expressed as the linear combination of signals from all $N_z$ redshift bins and the noise $n_i$,
\begin{equation}\label{E:elements}
I(\nu_i) = \sum_{j = 1}^{N_z} \widetilde{A}_{ij}\,N(z_j) + n_i,
\end{equation}
where $N(z_j)$ is the effective number of $\ell_*$ sources in redshift $z_j$ (Sec.~\ref{S:model_lightcone}), and $\widetilde{A}_{ij}$ converts $N(z_j)$ to the observed intensity in channel $\nu_i$. 

$\widetilde{A}$ is an $N_{\rm ch}\times N_{\rm z}$ matrix, where we have $N_{\rm ch}=70$ spectral channels, and we use $N_{\rm z}=2000$ ($\Delta z=5\times 10^{-4}$) redshift bins. In principle, we can set $\Delta z$ to infinitesimally small values;  however, in practice the redshift resolution is limited by the instrument spectral resolution, as the SLEDs of nearby galaxies can be highly degenerate when their respective spectral lines fall in the same set of observed channels. Setting $N_{\rm z}=2000$ in fact gives a much finer redshift resolution compared to the spectral channel width. We will use this fine redshift resolution as a starting point, and discuss the strategy to reduce redshift bins and therefore remove redundant information at the end of Sec.~\ref{S:methods_matrix}.

\begin{figure}[ht!]
\begin{center}
\includegraphics[width=\linewidth]{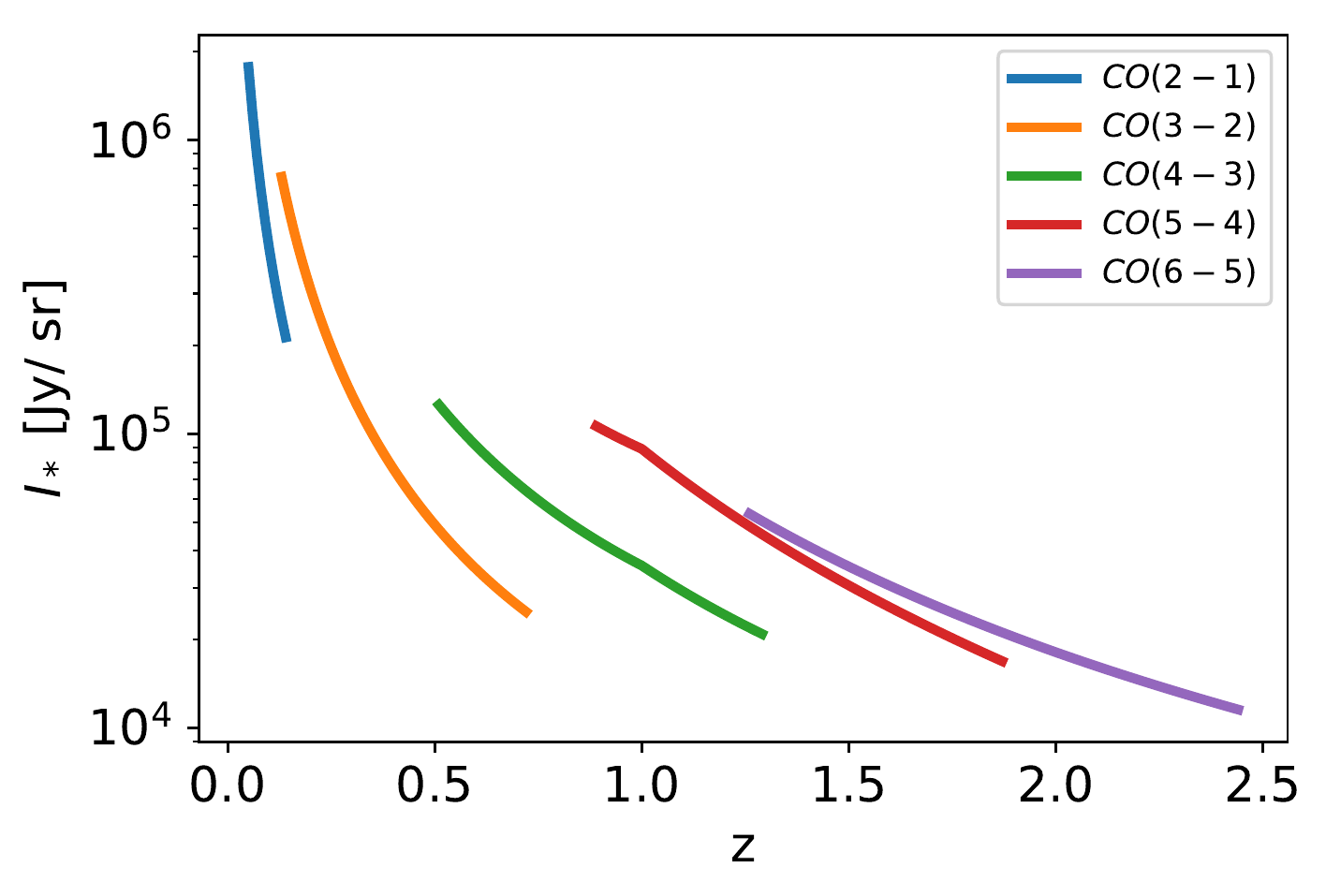}
\caption{\label{F:Istr}Intensity of the CO lines $I_*$ from the sources of characteristic luminosity $\ell_*$.}
\end{center}
\end{figure}

Most of the elements in $\widetilde{A}_{ij}$ are zeros except for which the sources at $z_j$ emit a spectral line at the observed frequency $\nu_i$. In this case,
\begin{equation}
\widetilde{A}_{ij} \equiv I_*^{\rm line}(z_j) = \ell_*^{\rm line}(z_j)\frac{1}{4\pi D_L^2(z_j)\delta\nu_i\Omega_{\rm pix}},
\end{equation}
where $\ell_*^{\rm line}$ is the line luminosity of an $\ell_*$ source in the model, $D_L$ is the luminosity distance, and $I_*^{\rm line}$ is defined as the observed line intensity of an $\ell_*^{\rm line}(z_j)$ source at $z_j$. Fig.~\ref{F:Istr} shows the $I_*^{\rm line}$ in our assumed model\footnote{The glitch in CO(5--4) at $z\sim 1$ is due to a feature in the Schechter luminosity function model in \cite{2016MNRAS.461...93P}. In their CO(5--4) luminosity function fit at $z=1$, a slightly higher $L_*$ and and a lower $\alpha$ value are derived that cause the slight apparent  discontinuity in redshift, even though the luminosity function does not show an abrupt change at $z=1$.}. Note that Eq.~\ref{E:elements} assumes that all of the sources at the same redshift have the same SLED, and we will first build our technique based on this assumption. In reality, the SLED varies across galaxy type. In Sec.~\ref{S:model_uncertainty}, we will show that our method also works in the realistic level of SLED variation.

\begin{figure*}[ht!]
\begin{center}
\includegraphics[width=\linewidth]{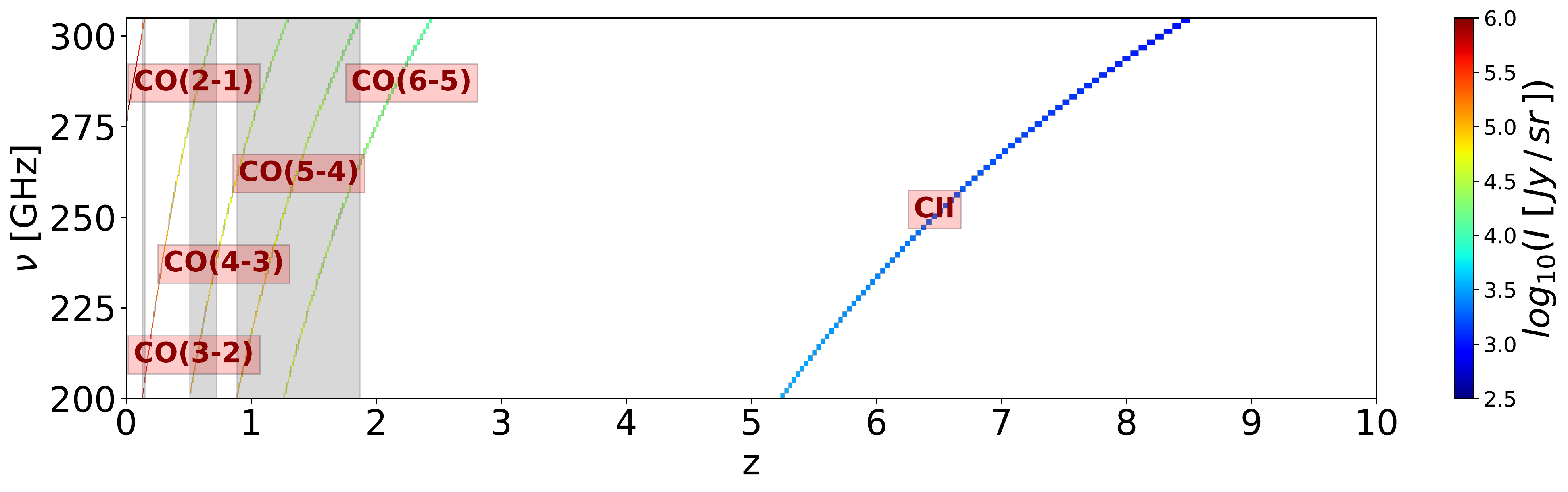}
\includegraphics[width=\linewidth]{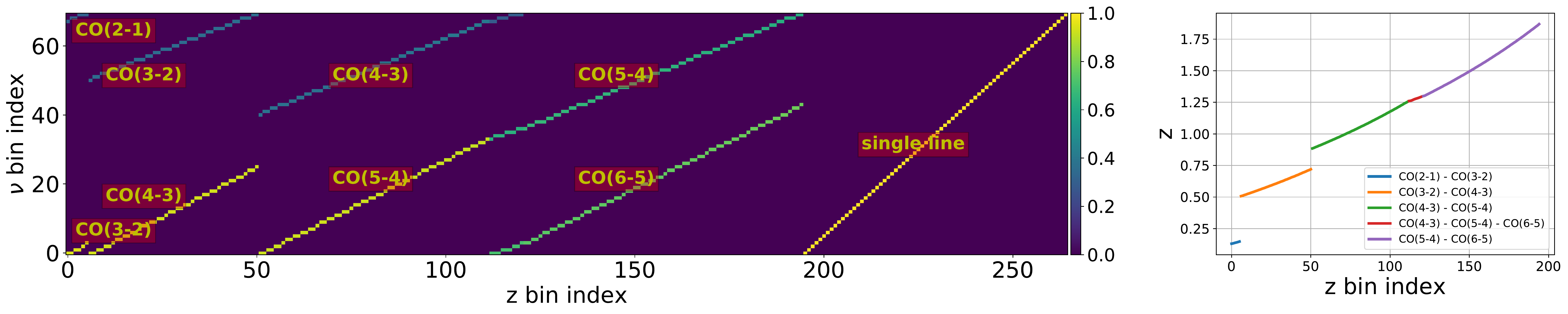}
\caption{\label{F:A} Top: $\widetilde{\mathbf{A}}$ matrix with our model on 70 frequency channels and 2000 redshift bins. $\widetilde{\mathbf{A}}$ are zeros (white) for the majority of the elements, and the six curves from left to right correspond to the six spectral lines, CO(2--1), CO(3--2), CO(4--3), CO(5--4), CO(6--5), and [\ion{C}{ii}]. The color scale indicates line intensities in the fiducial model. The gray shaded regions are the redshifts at which sources can be observed in multiple lines, and thus can be reconstructed in our technique. Bottom Left: $\mathbf{A}$ matrix with $N_{\rm ch}\times N_{\rm z} = 70 \times 265$ size, which is the reduced and normalized $\widetilde{\mathbf{A}}$. The color scale represents the intensities with the fiducial model SLED normalized within each column, i.e. for all column $j$, $\sum_{i=1}^{N_{\rm ch}}\mathbf{A}^2_{ij}=1$.,  Bottom Right: the redshift of the 195 multi-line redshift bins in $\mathbf{A}$, which are the redshift bins that have multiple CO lines observable in our mock survey. The colors label the pairs of detectable CO lines. The redshifts not covered by multiple CO lines cannot be reconstructed with our technique ($0.15\leq z\leq0.51$; $0.72\leq z\leq0.89$; $ z\geq1.87$).}
\end{center}
\end{figure*}

Eq.~\ref{E:elements} can be written in the matrix form,
\begin{equation}\label{E:matrixAraw}
\mathbf{I} = \widetilde{\mathbf{A}}\mathbf{N} + \mathbf{n},
\end{equation}
where $\mathbf{I}$ and $\mathbf{n}$ are $N_{\rm ch}$-element column vectors, $\mathbf{N}$ is an $N_z$-element column vector, and $\widetilde{\mathbf{A}}$ is an $N_{\rm ch}\times N_{\rm z}$ matrix. The top panel of Fig.~\ref{F:A} shows the $\widetilde{\mathbf{A}}$ matrix in our model. The $\widetilde{\mathbf{A}}$ matrix is mostly zeros, and the six curves from left to right  are the six spectral lines, CO(2--1), CO(3--2), CO(4--3), CO(5--4), CO(6--5), and [\ion{C}{ii}].

\begin{deluxetable*}{ccccccc}
\tablenum{1}
\tablecaption{\label{T:bands}Frequencies and Redshifts of the Six Defined Broad Bands}
\tablewidth{0pt}
\tablehead{
\colhead{Name} & \colhead{Line} & \colhead{$\nu$ bin index} & \colhead{$\nu$ (GHz)} &  \colhead{$\langle z \rangle (z_{\rm min} - z_{\rm max})$}
}
\startdata
J3 high    &CO(3--2)   & 51--69 (19 bins)   & 200.75--227.75   & 0.61 (0.51--0.72)\\ 
J4 low     &CO(4--3)   & 0--24 (25 bins)  & 268.25--304.25     & 0.61 (0.51--0.72)\\ 
J4 high    &CO(4--3)   & 41--69 (29 bins)   & 200.75--242.75   & 1.09 (0.89--1.30)\\ 
J5 low     &CO(5--4)   & 0--36 (37 bins)  & 250.25--304.25    & 1.09 (0.89--1.30)\\ 
J5 high    &CO(5--4)   & 37--69 (33 bins)   & 200.75--248.75   & 1.59 (1.30--1.87)\\ 
J6 low     &CO(6--5)   & 0--42 (43 bins)  & 241.25--304.25     & 1.56 (1.26--1.87)\\ 
\enddata
\tablecomments{The CO(2--1) and CO(3--2) overlapping redshifts of $0.13<z<0.5$ can also be reconstructed, but they are only covered by four frequency channels, which makes it difficult to quantify the reconstruction performance with sufficient statistical power. Therefore, we ignore these redshifts in our analysis.}
\end{deluxetable*}

The goal of our line de-confusion technique is to solve for the source count vector $\mathbf{N}$ in a given observed light cone data $\mathbf{I}$ and a model $\widetilde{\mathbf{A}}$ matrix. With the $\mathbf{N}$ solution, the intensity map of individual spectral lines $\mathbf{I}_{\rm line}$ can be reconstructed by $\mathbf{I}_{\rm line}= \widetilde{\mathbf{A}}_{\rm line}\mathbf{N}$, where $ \widetilde{\mathbf{A}}_{\rm line}$ is $\widetilde{\mathbf{A}}$   when only the target spectral line signals are turned on.

We further rewrite Eq.~\ref{E:elements} by normalization and reducing the nuisance information. First, we normalize the columns in $\widetilde{\mathbf{A}}$ and move the normalization factor $\mathbf{I}^{\rm norm}_j$ to their $\mathbf{N}$ element:
\begin{equation}\label{E:matrixA}
\mathbf{I} = \mathbf{A}\widetilde{\mathbf{N}} + \mathbf{n},
\end{equation}
where 
\begin{equation}
\begin{split}
\mathbf{A}_{ij} &=  \widetilde{\mathbf{A}}_{ij}\,/\, \mathbf{I}^{\rm norm}_j,\\
\widetilde{\mathbf{N}}_j &= \mathbf{N}_j\, \mathbf{I}^{\rm norm}_j,\\
\sum_{i=1}^{N_{\rm ch}}&\mathbf{A}^2_{ij}=1.
\end{split}
\end{equation}

Next, we will reduce the nuisance or redundant elements in Eq.~\ref{E:matrixA}. The columns of $\mathbf{A}$ are the basis spanning the observed data space. In constructing $\widetilde{\mathbf{A}}$, we simply design the columns to be equally spaced redshift bins as shown in the top panel of Fig.~\ref{F:A}. However, this natural basis is highly degenerate. To remove the nuisance information, we first discard the redshift bins that are zero vectors in $\widetilde{\mathbf{A}}$. The sources in these redshifts do not emit lines in observable frequencies, and thus no information can be used to constrain their $N(z)$.  Second, for the redshift bins containing only one CO line, their normalized columns in $\mathbf{A}$ are identical to other columns having a [\ion{C}{ii}] signal in the same frequency channel. In other words, given an observed data $\mathbf{I}$, we cannot distinguish the origin of the source with a single line emission. Therefore, we combine these identical columns into a single column.  In conclusion, we keep the columns in $\mathbf{A}$ with redshift bins that can be observed with multiple spectral lines, plus an identity matrix for the redshift bins that only have a single detectable line.

For all of the redshifts that can be observed in multiple lines, we design the size of the redshift bins based on the following two competing considerations. On the one hand, the redshift bins have to be small enough to faithfully represent the emitting source distribution. 
On the other hand, finer redshift bins give larger $\widetilde{\mathbf{N}}$ size,  and therefore more unknown parameters to be solved. The information can be compressed by combining some neighboring redshift bins, which are highly degenerate, since they have signals in the same channels with similar amplitudes. Therefore, we design the redshift bins using the following procedures: we (1) generate $\mathbf{A}$ with fine redshift bins ($\Delta z = 5\times 10 ^{-4}$ from $z = 0$ to 10)\footnote{Note that with these fine bins, the same frequency bin can map to multiple redshift bins instead of an one-to-one mapping.}, (2) keep the columns with multiple lines, (3) then identify the group of neighboring columns that have signals in the same sets of channels, and (4) keep the medium bin and discard the others. With this process, we get 195 non-degenerate columns. Hereafter, ``multi-line redshift bins'' refers to these 195 redshift bins that can be detected in multiple channels. Finally, we append the $N_{\rm ch}$-sized identity matrix (that account for  the single-line redshift bins) to these 195 columns to generate $\mathbf{A}$. The bottom left panel in Fig.~\ref{F:A} shows matrix $\mathbf{A}$ which has size $N_{\rm ch} \times N_z = 70\times 265$ (195 multi-line redshifts plus 70 columns of identity matrix). The bottom right panel shows the redshift of the 195 multi-line bins, and their color labels are the pairs of detectable CO lines. We also define six broad bands from these pairs of lines by binning groups of channels. Table~\ref{T:bands} lists the definition of the broad bands. Note that the CO(2--1) and CO(3--2) overlapping redshifts of $0.13<z<0.5$ can also be reconstructed, but they are only covered by four frequency channels, which makes it difficult to quantify the reconstruction performance with sufficient statistical power. Therefore, we ignore these redshifts in our analysis.

Our line de-confusion technique can only solve $\widetilde{\mathbf{N}}$ in the 195 multi-line redshift bins. The last 70 elements in $\widetilde{\mathbf{N}}$ that correspond to the identity matrix in $\mathbf{A}$ are nuisance parameters, since they represent degenerate ``single-line'' signals from different redshifts -- that is to say, we cannot reconstruct the signals in the single-line redshifts, which are the regions not covered by the lines in The bottom left panel of Fig.~\ref{F:A} ($0.15\leq z\leq0.51$; $0.72\leq z\leq0.89$; $ z\geq1.87$). In the following analysis, we will only focus on the reconstruction of $\widetilde{\mathbf{N}}$ in the 195 multi-line redshift bins.

\subsection{Sparse Approximation}\label{S:methods_sparse}
The key step in our de-confusion technique is to solve for $\widetilde{\mathbf{N}}$ in Eq.~\ref{E:matrixA}, given the observed spectrum $\mathbf{I}$ and model $\mathbf{A}$. This type of linear system has been extensively studied in the context of CMB map making, in which $\mathbf{I}$, $\mathbf{n}$, and $\widetilde{\mathbf{N}}$ in Eq.~\ref{E:matrixA} can be analogized to the time-ordered data, time-stream noise, and the pointing matrix, respectively. However, contrary to the map-making problem, our system is an ill-posed problem as there are more unknown variables ($N_z = 265$) than the input data points ($N_{\rm ch} = 70$). Thus the standard map-making algorithm \citep[e.g.,][]{2002PhRvD..65b2003S} cannot be applied.

In Eq.~\ref{E:matrixA}, the columns of $\mathbf{A}$ form a basis for $\mathbf{I}$, and the solution $\widetilde{\mathbf{N}}$ is the  linear combination coefficient. The columns of $\mathbf{A}$ form an over-complete basis, since the $\mathbf{A}$ matrix is only of rank $N_{\rm ch}$, and thus the solution $\widetilde{\mathbf{N}}$ is not unique. Indeed, for any given observed data $\mathbf{I}$, there are infinite $\widetilde{\mathbf{N}}$ that can perfectly fit the input.

Nevertheless, Eq.~\ref{E:matrixA} can be solved with the ``sparse'' condition, which means the preferred solution of $\widetilde{\mathbf{N}}$  is the one with a small number of nonzero elements. With this constraint, we can solve Eq.~\ref{E:matrixA} with the following well-defined optimization problem:
\begin{equation}\label{E:sparse_eqn}
\underset{\left \| \widetilde{\mathbf{N}} \right \|_0}{\mathrm{argmin}}\frac{1}{N_{\rm ch}}\left \| \mathbf{I} - \mathbf{A}\widetilde{\mathbf{N}} \right \|^2_2 < \epsilon^2,
\end{equation}
where the $\ell_0$-norm $\left \| \cdot  \right \|_0$ is the number of nonzero elements, and $\epsilon$ sets the threshold of error tolerance of the fit.

This type of problem is known as ``sparse approximation,'' which has been extensively studied in the context of signal processing and compressive sensing \citep{Candes:2006:RUP:2263435.2272020,Donoho:2006:CS:2263438.2272089}. The sparse approximation algorithms solve the sparse representation of the signal in a ``dictionary'' that is composed of a set of ``atoms,'' and represent the signal in the data in terms of the linear combination of a few atoms in the dictionary. In Eq.~\ref{E:sparse_eqn}, the dictionary is the matrix $\mathbf{A}$, and the atoms are the column vectors of $\mathbf{A}$.

The sparse approximation can only be applied if $k \equiv\left \| \widetilde{\mathbf{N}} \right \|_0 \ll N_{\rm ch}$. Note that the sparsity of the problem is quantify by $k/N_{\rm ch}$ but not $k/N_z$, since $k/N_z$ can always be designed to be arbitrarily small by choosing a large basis (fine redshift bins). However, the degree of freedom in the solution is restricted by the input size $N_{\rm ch}$, and thus the $k\ll N_{\rm ch}$ condition prohibits the algorithm from using more parameters than the input degree of freedom to over-fit the data.

In general, LIM light cones are not sparse, since there is always a large number of faint sources in the typical luminosity function (e.g. Schechter function), so all of the elements in $\widetilde{\mathbf{N}}$ are nonzero. However, as mentioned in Sec.~\ref{S:model_lightcone}, the light cone signals are dominated by only a few bright sources, and the intensity field can be well described by them. Consequently, the parameter $k$ in our problem can be quantified by the ``effective'' number of these bright sources per voxel that contribute most of the emission. Following  \cite{2019ApJ...877...86C}, we define the effective number $N_{\rm eff}$ as:
\begin{equation}
N_{\rm eff}(z) \equiv \frac{\left( V_{\rm vox} \, \int d\ell \, \Phi(\ell,z) \, \ell \right)^2}{V_{\rm vox} \, \int d\ell \, \Phi(\ell,z) \, \ell^2},
\end{equation}
where $V_{\rm vox}$ is the voxel size of the redshift bin. Note that $\Phi$ is the number of sources per luminosity per voxel, and therefore $N_{\rm eff}$ is dimensionless and is  
proportional to the voxel size. $N_{\rm eff}$ can be interpreted as the reciprocal of the effective shot noise in LIM, which is analogous to the $1/N$ shot noise in a galaxy power spectrum. If the luminosity function $\Phi$ follows the Schechter function form, then $N_{\rm eff}$ is (approximately) the number of sources brighter than $\ell_*$, which contributes the majority of the emission. 

\begin{figure}[ht!]
\begin{center}
\includegraphics[width=\linewidth]{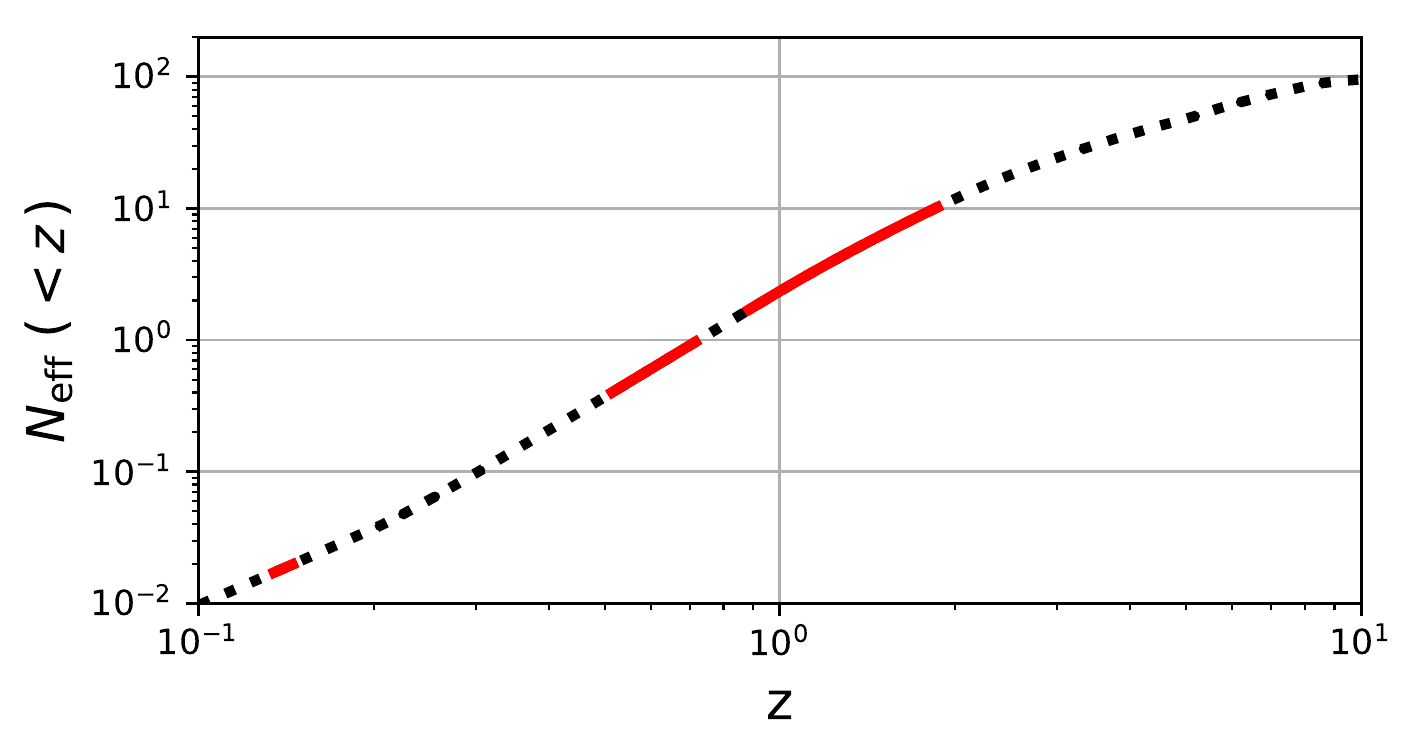}
\caption{\label{F:Neff} The redshift cumulated $N_{\rm eff}$ in our model (black dashed line). The red segments mark the multi-line redshift ranges that have multiple observable CO lines. Thus the de-confusion technique can be applied.}
\end{center}
\end{figure}

We can estimate $k$ by the cumulative $N_{\rm eff}(z)$ along the line of sight per light cone. Fig.~\ref{F:Neff} shows the cumulative $N_{\rm eff}$ in our model. While $N_{\rm eff}\sim 100$ from $z=0$ to 10, the only relevant range is $z\lesssim 2.5$, where CO lines fall in the observed frequency range (see Fig.~\ref{F:A}, top panel). Above $z\sim2.5$, only the high-$z$ [\ion{C}{ii}] lines can be observed, but they are much fainter then the CO signals and the assumed noise level, so they can be treated as background fluctuations. Therefore, for $z\lesssim2.5$, we find $N_{\rm eff}\sim 10\ll N_{\rm ch}=70$, so the sparse condition is qualified in our problem.

\subsection{The Matching Pursuit Algorithm}\label{S:methods_MP}
We use the matching pursuit (MP) algorithm first introduced by \citet{1993ITSP...41.3397M} to solve for Eq.~\ref{E:sparse_eqn}. The MP algorithm iteratively selects an ``atom'' in the ``dictionary'' to project out part of the signals in the data, and keep track of the current solution of the signal and residual for the next step until the stopping criteria is met. In our case, the column vectors in $\mathbf{A}$ are the atoms that form the dictionary space. A detailed description of the MP algorithm is in Appendix~\ref{A:MPA}.

In each step of the MP algorithm, the selected atom is the one that has the maximum inner product with the residual. The S/N of the signals in each step is the ratio of that maximum inner product to the instrument noise level $\sigma_n$ (see Appendix~\ref{A:proof_u_var} for the proof). Therefore, if we set the stopping criteria to be the maximum inner product smaller than $m$ times of $\sigma_n$, then this is an $m$-$\sigma$ detection threshold on the signals (e.g., $m=5$ for a 5$\sigma$ detection).  Note that the detection threshold here is based on the combined information in multiple spectral channels projected onto the dictionary space.

The choice of detection threshold ``$m$'' is a trade-off between the purity and completeness of the source extraction. Higher ``$m$'' values give a higher purity map, whereas lower ``$m$'' values pick out fainter sources at the cost of increased false detections from noise. The optimal value of ``$m$'' depends on the instrument sensitivity and the purpose of the reconstruction map. For example, to reconstruct the line luminosity function, one might use a higher threshold to reduce the false detections at the faint end; whereas to constrain the large-scale structure, a lower threshold is preferred to reduce the shot noise in the power spectrum. An analytical  formalism to determine the optimal threshold is to make use of the Fisher information framework \citep{2019ApJ...877...86C}, where one calculates the expectation value of the desired observable (e.g. power spectrum) as a function of threshold for a given signal model and noise level, and estimates the threshold that optimizes the Fisher information. Alternatively, one can simply perform test simulations with different thresholds to determine the optimal value. Both approaches can provide guidance on choosing the optimal threshold for the problem at hand, and a detailed investigation is beyond the scope of this paper.

\section{Results}\label{S:results}
We present line de-confusion results in the simple case where  mock light cones and the template ($\mathbf{A}$) are both generated from the same signal model (Sec.~\ref{S:model_lightcone}). We demonstrate that in this scenario, our technique is capable of extracting low-$z$ CO signals in the presence of realistic instrumental noise. We discuss the robustness of the performance against uncertainties in the signal model and contamination from astrophysical foregrounds, and extend the application to spectral lines in different wavelengths in Sec.~\ref{S:discussion}.

We quantify the reconstruction performance by computing two statistics on the true and reconstructed data: (1) the Pearson correlation coefficient (Sec.~\ref{S:Pearson}) and (2) the voxel intensity distribution (VID; Sec.~\ref{S:VID}). The former quantifies the phase-space information, whereas the latter captures the one-point statistics that describes the distribution of voxel intensities.

Finally, we present results with a variety of instrument noise levels $\sigma_n$ and the reconstruction threshold $m$. For each test, we use 2500 mock light cones to calculate the correlation coefficient and VID, and estimate errors with 100 noise realizations.

\subsection{Visualization of Example Results}
\begin{figure}[ht!]
\begin{center}
\includegraphics[width=\linewidth]{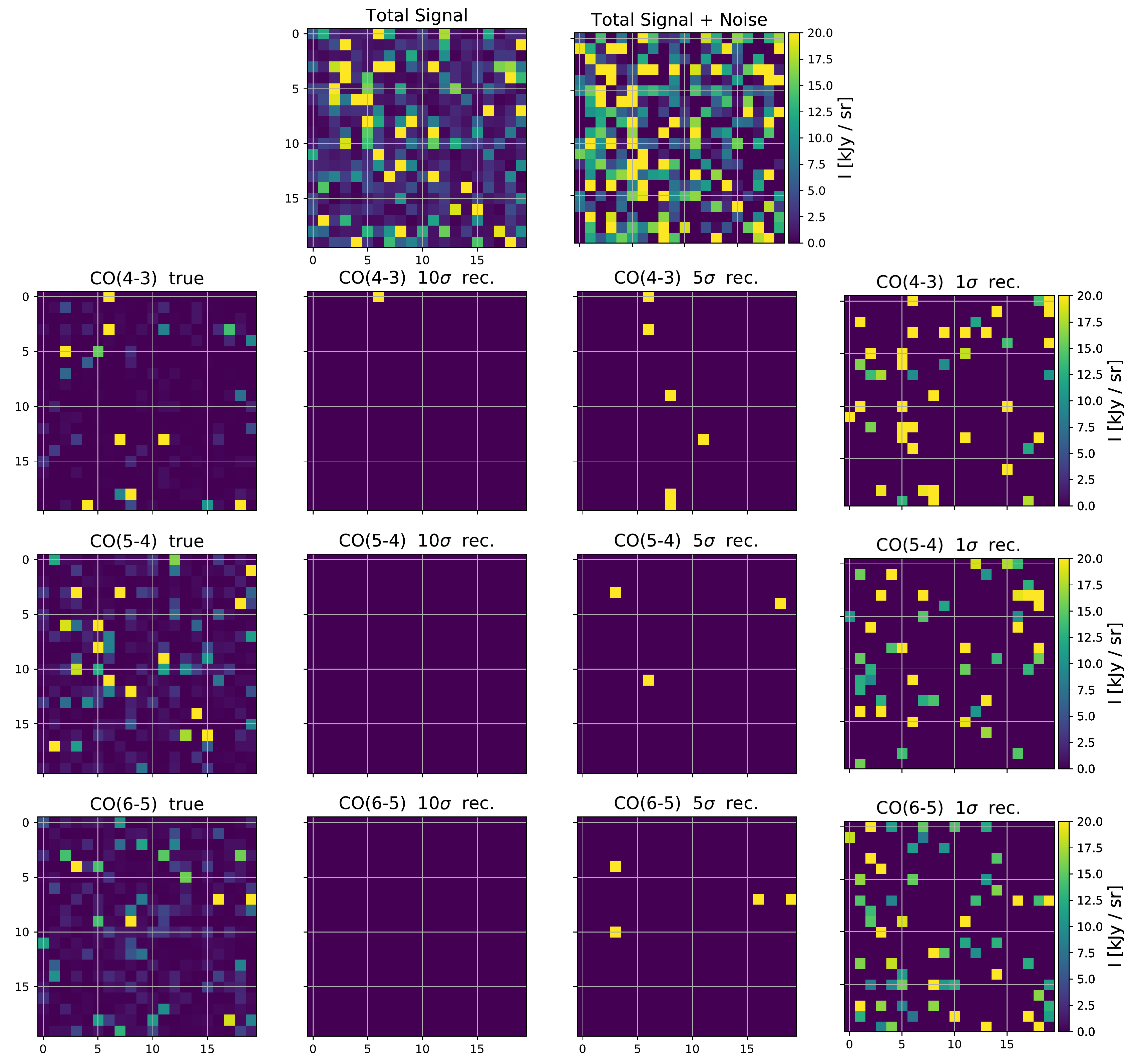}
\caption{\label{F:gen_im} Visualization of the phase-space reconstruction with $\sigma_n=10^4$ Jy sr$^{-1}$. We place the 400 toy model light cones into a $20\times 20$ pixel map, and show the reconstruction results on one of the spectrum bins (274 GHz). The top left panel is the true signal intensity map from all of the spectral lines (CO(3--2), CO(4--3), CO(5--4), CO(6--5), [\ion{C}{ii}]). The top right panel is the observed intensity map including line signals (top left panel) and noise. The bottom three panels show the true input and the 10$\sigma$, 5$\sigma$, and 1$\sigma$ reconstructed (right) emission field of the three spectral lines in the multi-line regime where the signal can be reconstructed.}
\end{center}
\end{figure}

Fig.~\ref{F:gen_im} visualizes the reconstruction results on one of the spectrum bins (274 GHz) that contains five lines (CO(3--2), CO(4--3), CO(5--4), CO(6--5), [\ion{C}{ii}]) in the observable band. We place the 400 light cones into a $20\times 20$ pixel map, and show the true and the reconstructed intensity fields. Note that the input signals do not exhibit the spatial clustering because each light cone is generated independently. The top left panel shows the total signal intensity from all five lines, and the top right panel shows the observed intensity including the total line signals and a Gaussian instrument noise with $\sigma_n=10^4$ Jy sr$^{-1}$. In this channel, three of the lines (CO(4--3), CO(5--4), CO(6--5)) are in the multi-line regime, so they can be reconstructed with our algorithm. The three bottom panels compare the true input to the reconstructed intensity maps for these lines with 10$\sigma$, 5$\sigma$, and 1$\sigma$ reconstruction threshold as the MP algorithm stopping criteria.

The choice of threshold is a trade-off between the completeness and the purity in the reconstructed map. As shown in Fig.~\ref{F:gen_im}, 
in the high threshold case (10$\sigma$), the MP algorithm only extracts a few bright sources that are above the reconstruction threshold. As the threshold decreases, more sources are being reconstructed, at the cost of increased false positive detection from the noise fluctuations or the interlopers. This example provides a visual depiction of the reconstruction algorithm. To further quantify the reconstruction performance, we consider two summary statistics in the following sections.

\subsection{Pearson Correlation Coefficient}\label{S:Pearson}
\begin{figure}[ht!]
\begin{center}
\includegraphics[width=\linewidth]{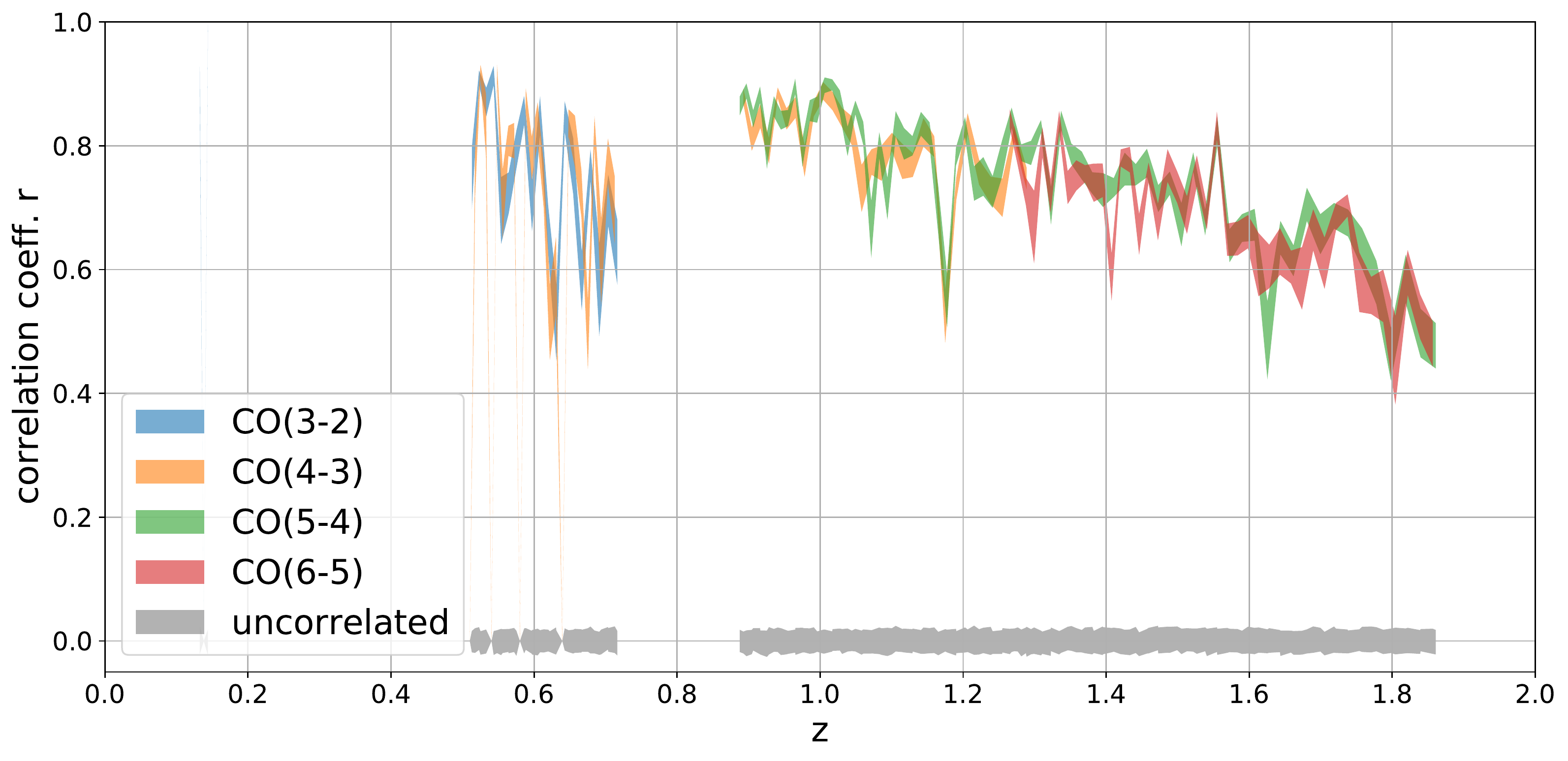}
\caption{\label{F:gen_r} Pearson correlation coefficient $r$ between the true and the reconstructed maps on 2500 light cones with $\sigma_n=10^4$ Jy sr$^{-1}$ 5$\sigma$ reconstruction. The bands are the 1$\sigma$ scatter of 100 noise realizations with the sample line signal. The gray bands are the $r$ value with the white noise map for reference.}
\end{center}
\end{figure}

We quantify the reconstruction performance by the Pearson correlation coefficient between the true and the reconstructed maps in each channel. The Pearson correlation coefficient is defined by
\begin{equation}\label{E:pearson}
r = \frac{\sum\limits_{i=1}^{N_{\rm lc}}\left ( I_{\rm true}^i - \left \langle I_{\rm true}  \right \rangle \right )\left ( I_{\rm rec}^i - \left \langle I_{\rm rec}  \right \rangle \right )}{\sqrt{\sum\limits_{i=1}^{N_{\rm lc}}\left ( I_{\rm true}^i - \left \langle I_{\rm true}  \right \rangle \right )^2}\sqrt{\sum\limits_{i=1}^{N_{\rm lc}}\left ( I_{\rm rec}^i - \left \langle I_{\rm rec}  \right \rangle \right )^2}},
\end{equation}
where $N_{\rm lc}=2500$ is the number of light cones, $I_{\rm true}^i$ and $I_{\rm rec}^i$ are the true and the reconstructed line intensity maps, respectively, at the $i$th light cone. Fig.~\ref{F:gen_r} shows the results of the correlation coefficient with a $\sigma_n = 10^4$ Jy sr$^{-1}$ noise level and a 5$\sigma$ reconstruction threshold. Our reconstructed map achieves $\sim 80\%$ correlation with the true input map at $z\lesssim 1.5$.

\begin{figure*}[ht!]
\begin{center}
\includegraphics[width=\linewidth]{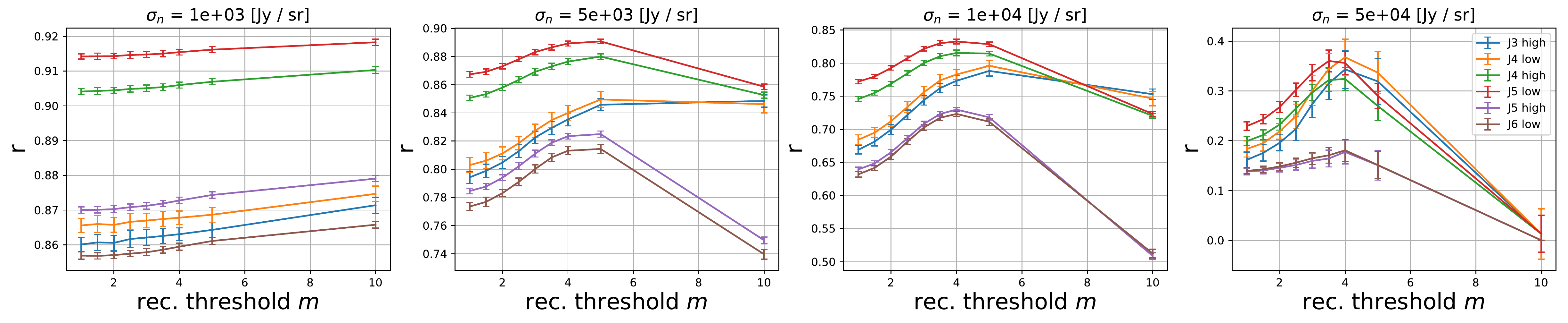}
\caption{\label{F:gen_r_bands} Average correlation coefficient $r$ in each broad band (Table~\ref{T:bands}) 
 within a range of noise level $\sigma_n$ and the reconstruction threshold $m$. The error bars are the rms of 100 noise realizations.}
\end{center}
\end{figure*}

Fig.~\ref{F:gen_r_bands} shows the correlation coefficient $r$ within a range of noise level $\sigma_n$ and the reconstruction threshold $m$. For simplicity, we show the average $r$ value of each broad band defined in Table~\ref{T:bands}. The key findings are summarized below.

\begin{itemize}
\item At a fixed $m$ value, $r$ decreases as $\sigma_n$ increases. This is because under the same purity (same $m$), the detection threshold $m\sigma_n$ is higher for higher $\sigma_n$, and thus fewer sources have been reconstructed.
\item The six defined bands correspond to three pairs of lines from different redshift bands (Table~\ref{T:bands}). The lines within each pair are strongly correlated because they are the signals from the same sources, and thus they are reconstructed in the same MP iteration. The pairs of lines in the same redshift are reconstructed in the same MP iteration, and as a result, they are highly correlated. 
\item Because of the purity and completeness trade-off, the maximum correlation $r$ happens at the intermediate threshold $m$ (except for the lowest-noise $\sigma_n=10^3$ Jy sr$^{-1}$ case, discussed in the next enumerated point). 
\item Correlation coefficient $r$ has very low dependency on $\sigma_{\rm th}$ in the $\sigma_n=10^3$ Jy sr$^{-1}$ case. This can be understood by comparing the noise level $\sigma_n$ to the quantity $I_*$, the intensity of the $\ell_*$ source in the Schechter function (Fig.~\ref{F:Istr}). At the redshift range in which we perform the reconstruction ($0.5\lesssim z\lesssim 1.9$), $10^4\lesssim I_*\lesssim 10^5$ Jy sr$^{-1}$, which indicates that $m\sigma_n < I_*$ for all of the $m$ values considered in Fig.~\ref{F:gen_r_bands} ($m=1\sim10$). In the Schechter luminosity function, the sources $\gtrsim \ell_*$ contribute the majority of the information in the intensity field \citep{2019ApJ...877...86C}, and thus the correlation coefficient $r$ is not sensitive to the change in reconstruction threshold if $m\sigma_n\ll\ell_*$. 
\end{itemize}

\subsection{VID}\label{S:VID}

\begin{figure*}[ht!]
\begin{center}
\includegraphics[width=\linewidth]{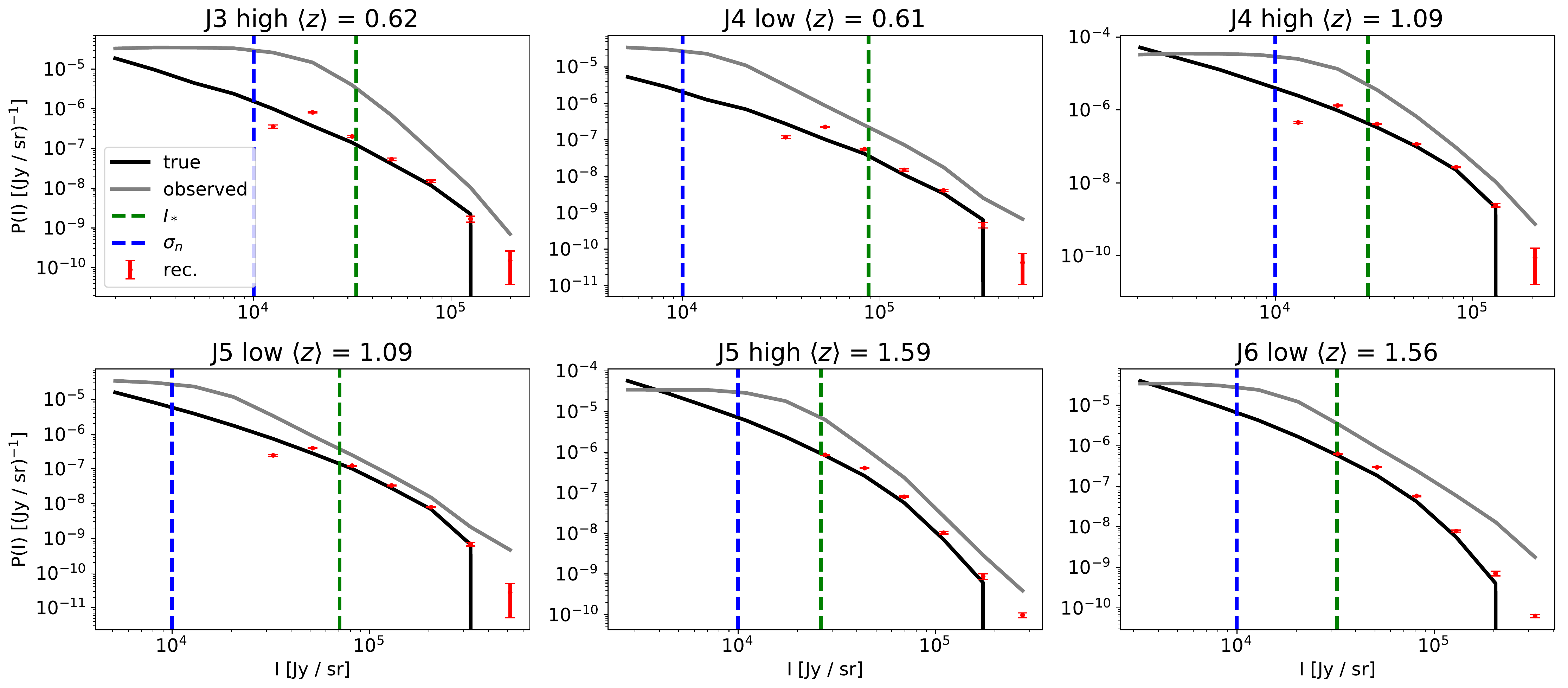}
\caption{\label{F:gen_VID} VID of the true input (black) and the reconstructed (red) maps in the six broad bands with noise level $\sigma_n=10^4$ Jy sr$^{-1}$ and reconstruction threshold $m=4$.  The error bars are the rms of 100 noise realizations. The gray curves are the VID of the total observed map, which includes signals from all of the lines and noise. For reference, the blue and green dashed lines mark the noise level $\sigma_n$ and $I_*$, respectively.}
\end{center}
\end{figure*}

The Pearson correlation coefficient traces the phase-space variations between the true and reconstructed maps, but it cannot distinguish a systematic constant offset, i.e., if the reconstructed line signals are systematically lower or higher than the true input. Therefore, we check the consistency of the reconstructed and true input maps using the one-point statistics, VID. Note that we do not directly compare the mean intensity of each map since we only reconstruct the bright sources that are above the threshold and neglect all faint sources in the reconstructed map, so the mean intensity is not expected be faithfully recovered.

The VID of an LIM map contains information beyond the power spectrum, and is valuable for LIM targeting a late-time universe where the large-scale structure is highly non-Gaussian and cannot be fully described by two-point statistics. For example, \citet{2017MNRAS.467.2996B} showed that the VID can constrain the luminosity function model parameters, and \citet{2019ApJ...871...75I} demonstrated that a joint analysis of the power spectrum and VID improved the constraining power on the source luminosity function. 

Fig.~\ref{F:gen_VID} compares the VID of the true and reconstructed maps in the six broad bands with noise level $\sigma_n=10^4$ Jy  sr$^{-1}$ and reconstruction threshold $m=4$. The gray lines compose the VID of the total observed map, which includes signals from all of the lines and noise. The black lines are the VID of the target line maps, and the red data points are the VID of the reconstructed target line map.While the VID of the total observed map is one to two orders of magnitude above that of the target line signal, our reconstruction technique can faithfully recover the VID of the signal to slightly below the $I_*$-scale, the characteristic source luminosity in the Schechter function.

In this realistic survey setup with the assumed signal model, we show that our method can successfully reconstruct the VID of the CO signal down to $\sim \ell_*$ scales. This provides a strong constraint on the CO luminosity function, as both are 1D statistics of the intensity field and are closely related. The CO luminosity function at various redshifts provides valuable insight on the formation and evolution of galaxies  across cosmic time. Specifically, CO is a tracer of H$_2$ gas in the interstellar medium, and can therefore be used to study the evolution of the molecular gas content and their distribution as a function of time \citep{2014ApJ...782...79W,2019ApJ...882..138D,2019ApJ...872....7R}. We note though that the expected S/N on the luminosity function depends on the assumed model.

\section{Discussion}\label{S:discussion}
\subsection{Model Uncertainty}\label{S:model_uncertainty}
\begin{figure*}[h!]
\begin{center}
\includegraphics[width=\linewidth]{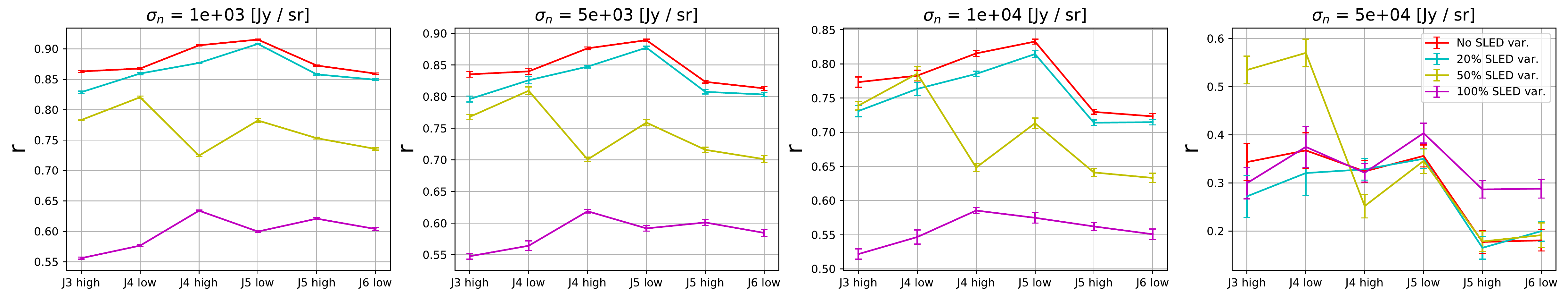}
\caption{\label{F:gen_r_bands_var} Comparing $r$ of the no SLED variation (red), 20\% variation (cyan), 50\% variation (yellow), and 100\% variation (purple) with the 4$\sigma$ reconstruction threshold. The values are the average of $r$ within the channels of the band, and the error bars are the rms of 100 noise realizations of all of the spectral bins in each band.}
\includegraphics[width=\linewidth]{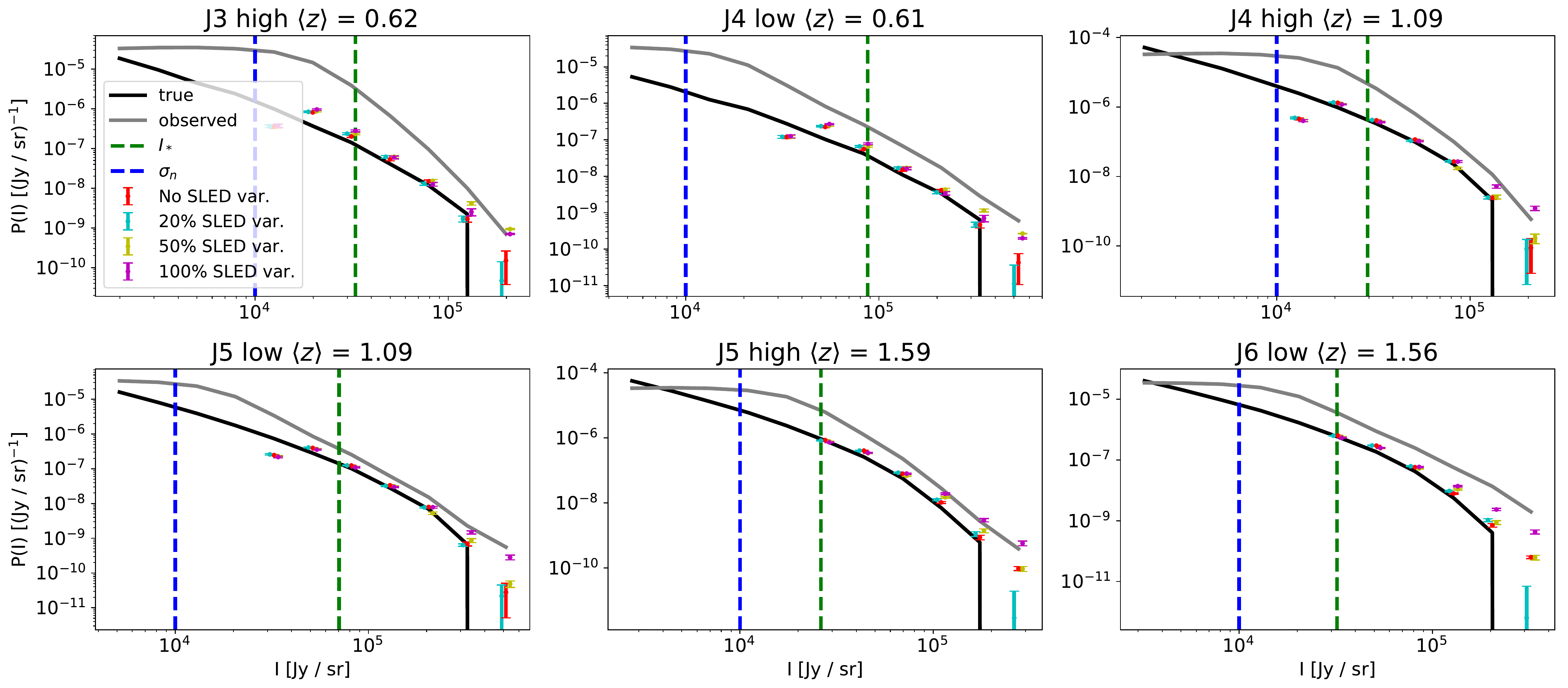}
\caption{\label{F:gen_VID_var} VID of the no SLED variation (red), 20\% variation (cyan), 50\% variation (yellow), and 100\% variation (purple), and the true (black) maps in the six broad bands with noise level $\sigma_n=10^4$ Jy sr$^{-1}$ and reconstruction threshold $m=4$.  The error bars are the r.m.s. of 100 noise realizations. The gray curves are the VID of the total observed map, which includes signals from all of the lines and noise. For reference, the blue and green dashed lines mark the noise level $\sigma_n$ and $I_*$.}
\end{center}
\end{figure*}

For the results presented in Sec.~\ref{S:results}, the light cone signals and the dictionary template $\mathbf{A}$ are both generated from the same assumed signal model (Sec.\ref{S: model}). However, in reality, the variation in SLED across galaxies will affect the reconstruction performance. To test how the SLED uncertainties affect the reconstruction, we apply three different SLED model variations and bias levels (at $20\%$, $50\%$, and $100\%$) to the mock data, and run the reconstruction with the same dictionary template $\mathbf{A}$. We detail the definition of variation and bias below. \citet{2015A&A...577A..46D} measured  multiple CO lines of ULIRGs at $z\sim 1.5$, and estimated a $\sim 20\%$ variation on the CO SLED ratio for their sample. Therefore, an assumed $50\%$ or $100\%$ variation can be more extreme than realistic variations.

\subsubsection{SLED model variation}

\begin{figure*}[ht!]
\begin{center}
\includegraphics[width=\linewidth]{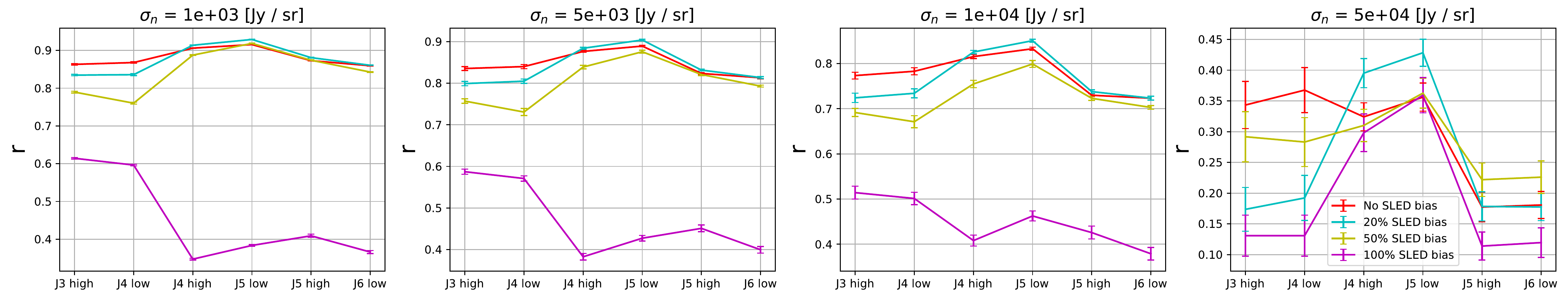}
\caption{\label{F:gen_r_bands_bias} Comparison $r$ of the no SLED bias (red), 20\% bias (cyan), 50\% bias (yellow), and 100\% bias (purple) with a 4$\sigma$ reconstruction threshold. The values are the average of $r$ within the channels of the band, and the error bars are the rms of 100 noise realizations of all of the spectral bins in each band.}
\includegraphics[width=\linewidth]{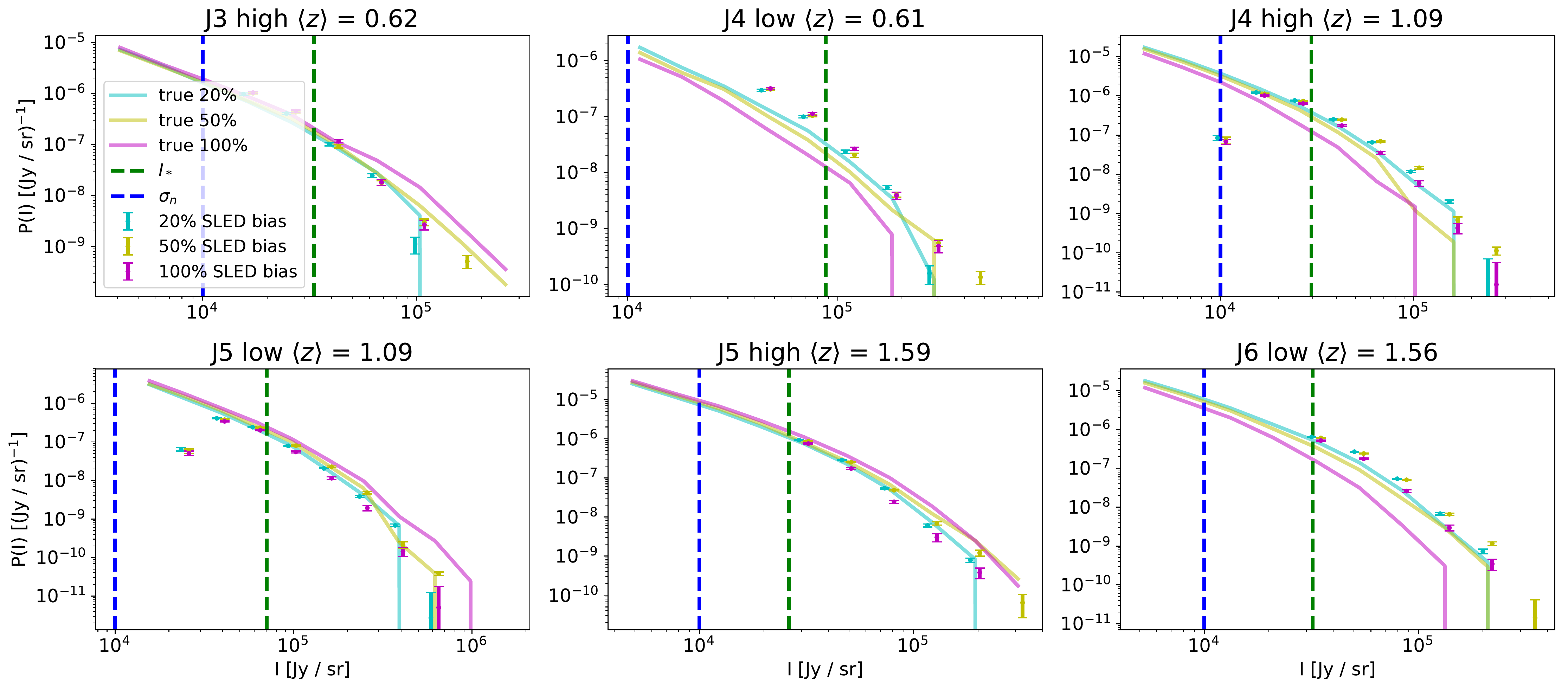}
\caption{\label{F:gen_VID_bias} VID of no SLED bias (red), 20\% bias (cyan), 50\% bias (yellow), 100\% bias (purple), and the true maps (solid lines) in the six broad band with noise level $\sigma_n=10^4$ Jy /sr and reconstruction threshold $m=4$.  The error bars are the r.m.s. of 100 noise realizations. The grey curves are the VID of the total observed map that includes signals from all the lines and noise. For reference, the blue and green dashed lines mark the noise level $\sigma_n$ and $I_*$.}
\end{center}
\end{figure*}

First, we test the case with SLED model variations. For each line of each source in the mock light cone, we assign a line luminosity $L^{\rm line}$:
\begin{equation}\label{E:SLED_var}
L^{\rm line} = L^{\rm line}_{\rm fid} (1+\delta_L),
\end{equation}
where $L^{\rm line}_{fid}$ is the fiducial luminosity from our model, and $\delta_L$ is a zero-mean Gaussian random variable with the standard deviations of $0.2$, $0.5$, and $1.0$ for the $20\%$, $50\%$, and $100\%$ variation cases, respectively.

Fig.~\ref{F:gen_r_bands_var} shows the reconstructed correlation coefficients $r$ with a 4$\sigma$ reconstruction threshold ($m=4$) with different SLED variations. We see that introducing $20\%$ SLED variations in the model has a very mild impact on the reconstructed correlation; whereas with 50\% (100\%) variation, the correlation coefficient drops by about 10\% ($\sim 50\%$). Fig.~\ref{F:gen_VID_var} compares the VID. We see that the SLED fluctuation has a negligible impact on the VID reconstruction in the 20\%, 50\%, and 100\% variation cases. Hence, we conclude that our technique is robust against realistic level of CO SLED fluctuations.

\subsubsection{Model offset}

In addition to SLED variation, we also test whether our model gives biased estimates of the average SLED by assigning line luminosity $L^{\rm line}$ as
\begin{equation}
L^{\rm line} = L^{\rm line}_{\rm fid} (1+b_L),
\end{equation}
where $b_L$ is a constant offset that we assign to the model lines, and it is applied to all of the sources in the mock light cones. For the 20\% bias level, we apply $b_L=(+0.1,-0.1,+0.1,-0.1)$ for the four CO lines CO(3--2), CO(4--3), CO(5--4), and CO(6--5) to ensure the SLED ratio between neighboring lines is $~20\%$, since our algorithm is only sensitive to the line ratio between two neighboring CO lines.

Fig.~\ref{F:gen_r_bands_bias} and Fig.~\ref{F:gen_VID_bias} show the reconstructed correlation coefficients $r$ and VID with a 4$\sigma$ with three SLED bias levels. Similar to Fig.~\ref{F:gen_r_bands_var}, the correlation only drops significantly when the bias is tuned to 100\%. Therefore, our technique is also robust against a realistic level of potential CO SLED bias.

\subsection{Application to LIM in Other Wavelengths}
The technique developed in this work is not restricted to the [\ion{C}{ii}] and CO lines' blending problem. It can in principle be applied to a range of LIM experimental setups. As a demonstration, we apply our method to reconstruct near-infrared lines in an SPHEREx-like survey.

SPHEREx is an ongoing NASA MIDEX mission to conduct an all-sky near-infrared spectro-imaging survey \citep{2014arXiv1412.4872D}.\footnote{\url{http://spherex.caltech.edu}} SPHEREx will carry out the first all-sky spectral survey at wavelengths between 0.75 and 5 $\mu$m with 96 spectral channels and a ${6{''}.2}$ pixel size. Ly$\alpha$ (121.6 nm), H$\alpha$ (656.3 nm), H$\beta$ (486.1 nm), [\ion{O}{ii}] (372.7 nm), and [\ion{O}{iii}] (500.7 nm) are the five prominent lines detectable by SPHEREx across a range of redshifts especially in the LIM regime.

\begin{figure}[ht!]
\begin{center}
\includegraphics[width=\linewidth]{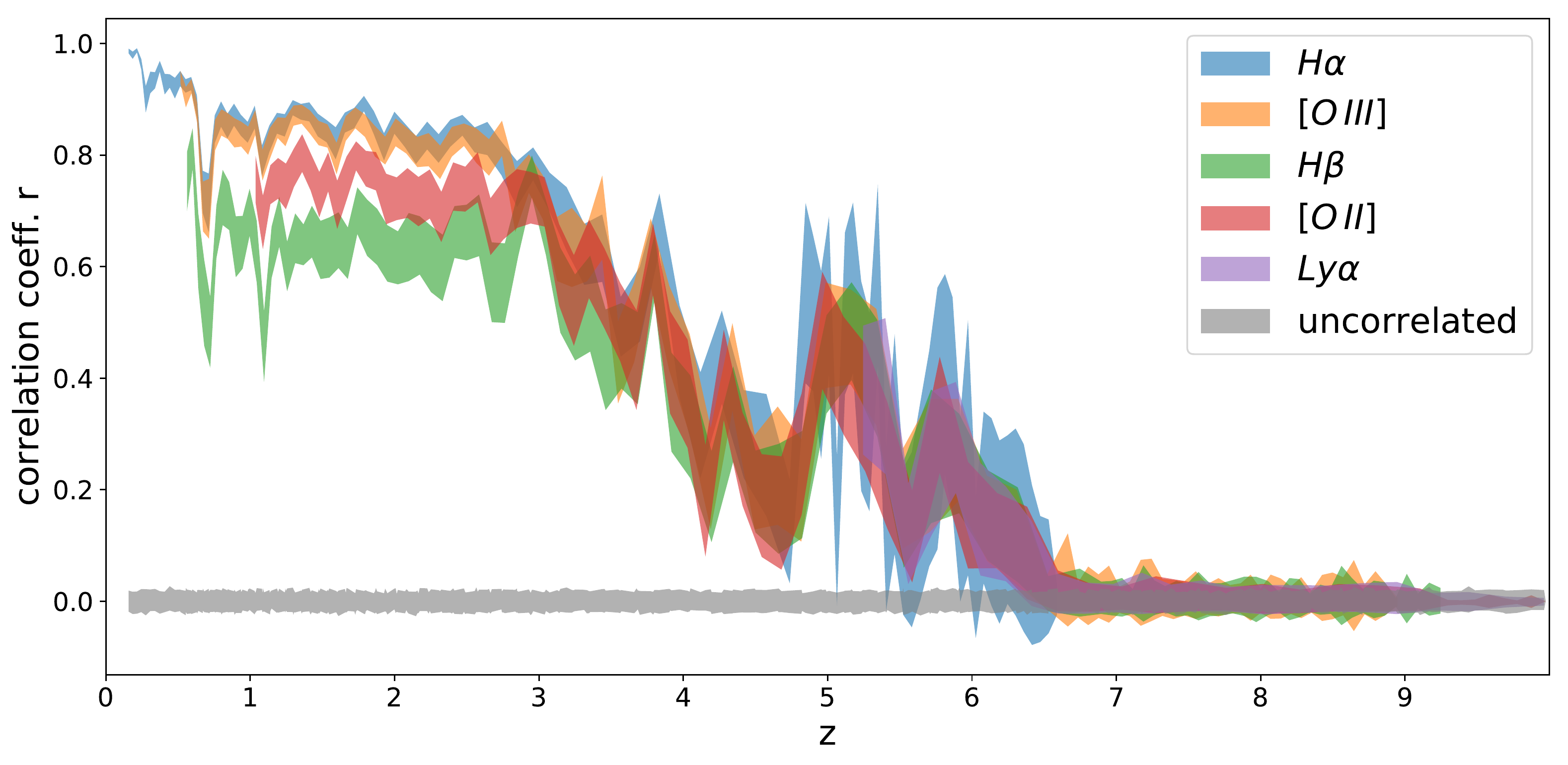}
\caption{\label{F:gen_r_spherex} Pearson correlation coefficient $r$ between the true and the reconstructed maps on 2500 light cones for SPHEREx-like mock data. The bands are the rms. of the value in 100 noise realizations with the sample line signal. The gray bands are the correlation coefficient with the uncorrelated white noise map for reference.}
\end{center}
\end{figure}

The line signal model is described in Appendix \ref{A:SPHEREx}. We generate a near-infrared LIM mock data with a ${6{''}.2}\times {6{''}.2}$ pixel size and a  $5\sigma$ point-source sensitivity of $m_{\rm AB}=22$ (similar depth as the SPHEREx deep fields), and run our de-confusion algorithm on the mock light cones. Fig.~\ref{F:gen_r_spherex} shows the results of the correlation coefficients between the true and a 3$\sigma$-threshold reconstructed intensity maps. The reconstructed map achieves $\sim 80\%$ correlation with the true input map at $z\lesssim 3$, and decreases toward higher redshifts, as the $I_*$ of the lines approaches the noise level $\sigma_n$. At $z\sim 5$, the brightest line, H$\alpha$, is redshifted into SPHEREx bands with a high spectral resolution of $R\sim 130$ and suffers less signal dilation, resulting in an increase of S/N on a single source detection, and thus $r$ slightly rebounds at this redshift.

\begin{figure}[ht!]
\begin{center}
\includegraphics[width=\linewidth]{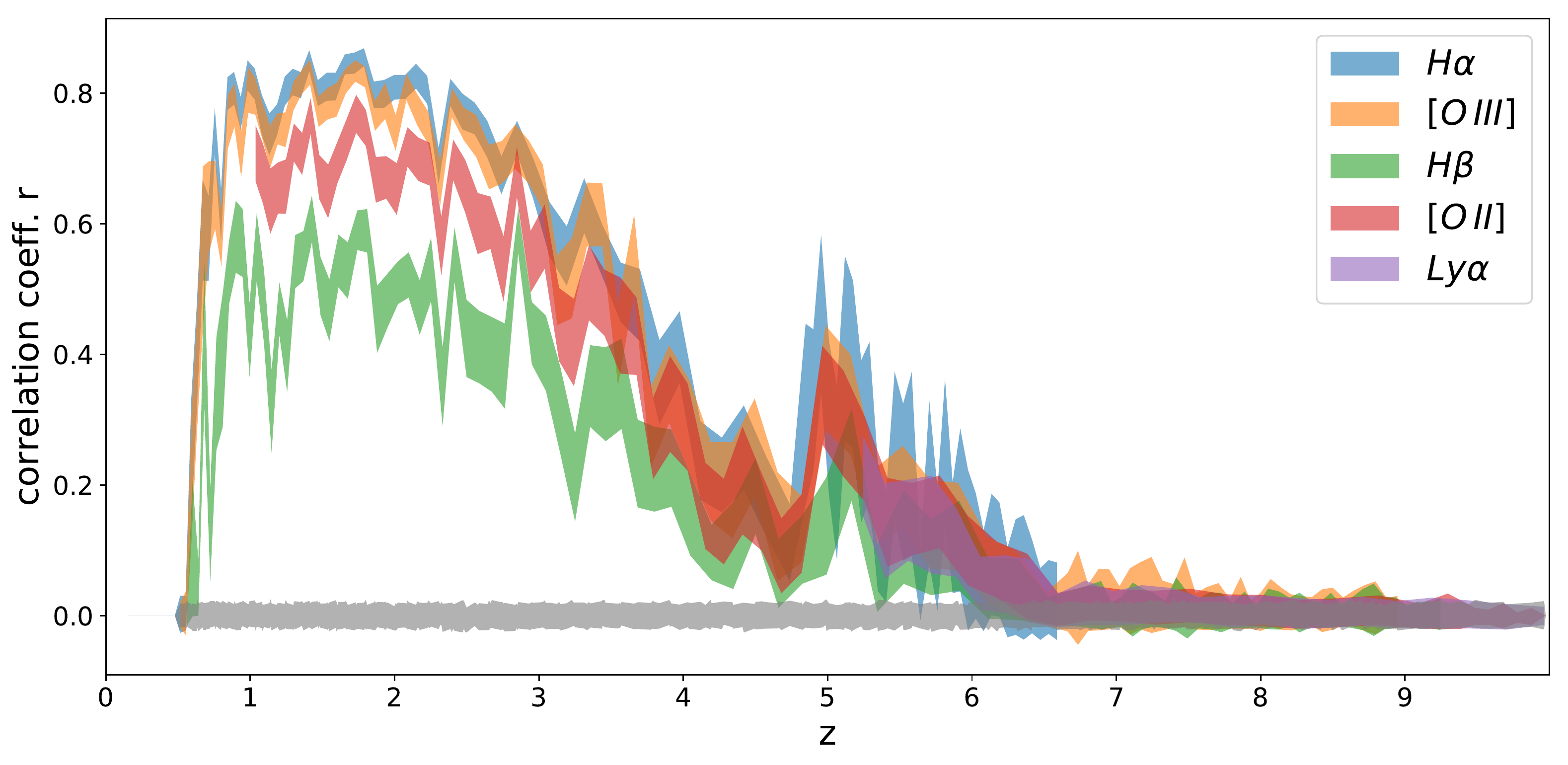}
\caption{\label{F:gen_r_spherex_var} Pearson correlation coefficient $r$ between the true and the reconstructed maps on 2500 light cones for SPHEREx-like mock data with a realistic level of SLED variation. The bands are the rms of the value in 100 noise realizations with the sample line signal. The gray bands are the correlation coefficient with the uncorrelated white noise map for reference.}
\end{center}
\end{figure}

To account for uncertainties in modeling the SLED, we apply a realistic level of line luminosity variations from \citet{2006ApJ...642..775M}. We apply 10\%, 50\%, and 100\% SLED variations (Eq.\ref{E:SLED_var}) to the line luminosity ratio of H$_\alpha$/H$\beta$, [\ion{O}{ii}]/H$\alpha$, and [\ion{O}{iii}]/[\ion{O}{ii}], respectively. Fig.~\ref{F:gen_r_spherex_var} shows the correlation coefficients between the true and a 3$\sigma$-threshold reconstructed intensity maps. Comparing with the fixed SLED case (Fig.~\ref{F:gen_r_spherex}), only the $H\beta$ line shows a significant decrease in the performance. The intensity map of the three brighter lines (H$\alpha$, [\ion{O}{ii}], and [\ion{O}{iii}]) can still be extracted with $\gtrsim 70\%$ correlation compared to the true input.

We conclude that our algorithm can reasonably well reconstruct the phase-space LIM signal in a SPHEREx-like experiment, given the expected variation of (redshifted) optical line ratios. The technique can be generalized to  different LIM experimental applications, and the reconstructions are fairly robust against uncertainties in the SLED modeling.

\subsection{Foreground Subtraction}\label{S:bg}
\begin{figure*}[ht!]
\begin{center}
\includegraphics[width=\linewidth]{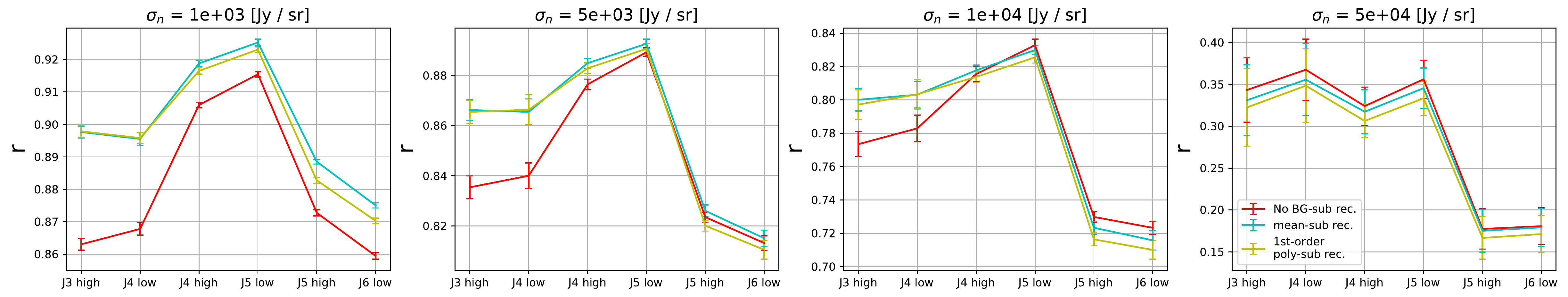}
\caption{\label{F:gen_r_bands_bg}Comparing $r$ of the reconstructed map with the no background subtraction (red), subtracting the mean of the whole data cube (cyan), and subtracting a first-order polynomial in the spectral direction for each light cone (yellow), using a 4$\sigma$ reconstruction threshold. The values are the average of $r$ within the channels of the band, and the error bars are the rms of 100 noise realizations of all of the spectral bins in each band.}
\includegraphics[width=\linewidth]{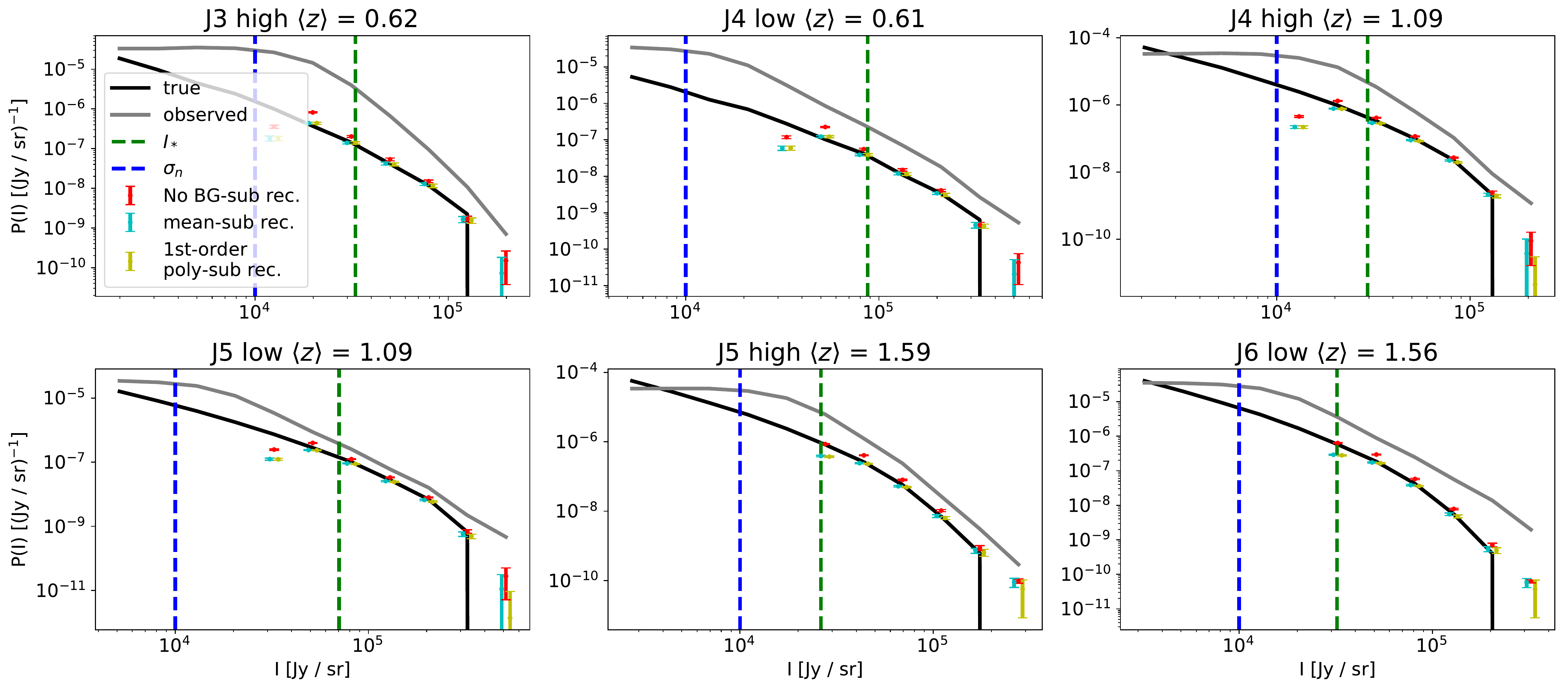}
\caption{\label{F:gen_VID_bg} VID of no background subtraction (red),subtracting the mean of the whole data cube (cyan), subtracting a first order polynomial in spectral direction for each light cone (yellow), and the true (black) maps in the six broad band with noise level $\sigma_n=10^4$ Jy /sr and reconstruction threshold $m=4$.  The error bars are the r.m.s. of 100 noise realizations. The grey curves are the VID of the total observed map that includes signals from all the lines and noise. For reference, the blue and green dashed lines mark the noise level $\sigma_n$ and $I_*$.}
\end{center}
\end{figure*}

In addition to line interlopers, LIM data are subjected to strong continuum foreground from various sources. For the frequency range considered in this work ($\sim$ 200-300 GHz), the dominant foregrounds are the atmospheric emission, dust continuum and the CMB; whereas for LIM in the near infrared, e.g. SPHEREx \citep{2014arXiv1412.4872D}, the zodiacal light and the galaxy stellar continuum are the dominant  continuum foregrounds. Even though these  foregrounds are brighter than the sought-after line signals, their spectral responses are expected to be smooth and are distinct from the spectral line features, so that the continuum foregrounds can be separated and mitigated, for example by a smooth function fit such as a low order polynomial\footnote{Some foreground components are also spatially smooth (e.g. zodiacal light) that can be filtered in the spatial domain as well.}. Here we test how the foreground mitigation process affects our line reconstruction results.

We consider two cases of foreground mitigation. First we emulate the foreground removal process in the presence of an approximately constant foreground in both the spatial and spectral dimensions, for example the zodiacal light. In this case, before running the reconstruction, we subtract the mean value of the whole data cube, i.e. the mean intensity in $N_{\rm lc}\times N_{\nu}$ voxels. The second case is to emulate the continuum subtraction process of the galaxy stellar or dust continuum, which are expected to have smooth spectra but different in each light cone, since each light cone contain different galaxies with different continuum spectrum. We fit and subtract a first order polynomial function to the spectrum of each light cone before running the reconstruction.

The results of a 4-$\sigma$-threshold reconstruction with different noise $\sigma_n$ level are shown in Fig.~\ref{F:gen_r_bands_bg} and Fig.~\ref{F:gen_VID_bg}. We can see compared to the no background subtraction case, the $r$ value is even higher in these two tests. This is because our reconstruction only extract the bright lines, and the fainter lines act as a background for the MP algorithm. The signals from the fainter lines introduce not only fluctuations but also a bias in the data, since the line signals are always positive unlike the zero-mean noise. The reconstruction performance is improved after background subtraction because this bias level is also removed during this process. For the VID results, we see that there is no significant difference compared to the no background subtraction case. This is again due to the fact that the background level is much fainter than the brightness of the sources being extracted with our algorithm, so the background subtraction have no impact on the reconstruction. In conclusion, the background subtraction in the LIM data reduction pipeline will not affect our line reconstruction technique.

\subsection{Prior with External Catalogs}\label{S:ext_cats}
\begin{figure*}[ht!]
\includegraphics[width=\linewidth]{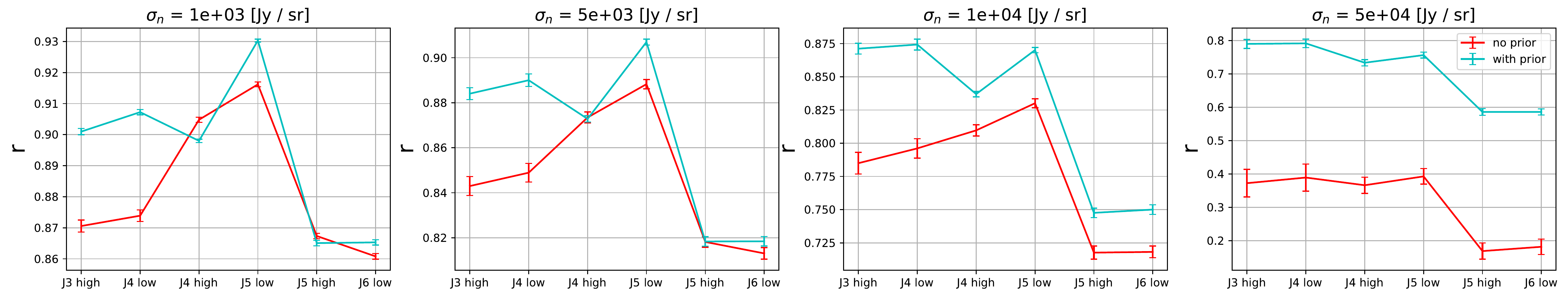}
\caption{\label{F:gen_r_bands_prior}Comparing $r$ of the reconstructed map with no external catalog prior (red), and utilizing the external catalog prior to fit the catalog sources before running the MP reconstruction (cyan), using 4-$\sigma$ reconstruction threshold. The values are the average of $r$ within the channels of the band, and the error bars are the r.m.s. of 100 noise realizations of all the spectral bins in each band.}
\end{figure*}

Our analysis uses the LIM data itself without invoking any external information. In practice, initial LIM survey fields are designed to in part overlap with existing photometric or spectroscopic galaxy surveys, and thus there will be information provided by external galaxy catalogs to aid the line de-confusion problem. Ignoring redshift uncertainties of the external catalogs, one simple approach to incorporate the external information is to force the MP algorithm to first select the redshift bins that contain galaxies from the catalog. After  iterating through the catalog sources, we then continue the normal MP procedure until hitting the stopping criteria. 

To test the effect of including prior knowledge from external catalogs, we generate a mock catalog by selecting sources with CO(5--4) flux greater than $150\, L_\odot$ Mpc$^{-2}$ ($6\times 10^{-17}$ W m$^{-2}$) in each light cone. The source density in the catalog is $\sim 1.2$ per light cone (integrated along line of sight) for a $0.43^2$ arcmin$^2$ pixel solid angle. The flux cut corresponds to an $L_*$ galaxy at $z\sim 2$. According to \citet{2012ApJ...752..113H}, such a $L_*$ galaxy has an absolute magnitude $M_{\rm AB}\sim -23$ in the optical, which gives an apparent magnitude of $m_{\rm AB}\sim 21.8$, approximately the depth of the assumed optical catalog.

With this mock external catalog, we identify the redshift bins containing the catalog sources, regardless of the noise level and threshold value. After projecting out these components, we run the MP algorithm on the residual data as per usual until we hit the stopping criteria. The results of a 4-$\sigma$-threshold reconstruction with different noise levels are shown in Fig.~\ref{F:gen_r_bands_prior}. The reconstruction shows improved results  for all noise levels. However, the huge improvement in the highest-noise case cannot be interpreted as a successful reconstruction of source intensities. At this high noise level, the signals are well buried under the noise, and when we fit the data with the catalog source redshift templates, the extracted components are dominated by noise rather than signal amplitude. Thus the improved correlation is merely due to the position information imposed by the external catalogs.

\subsection{Comparing with the Limit of No Interlopers}
\begin{figure}[htbp!]
\begin{center}
\includegraphics[width=\linewidth]{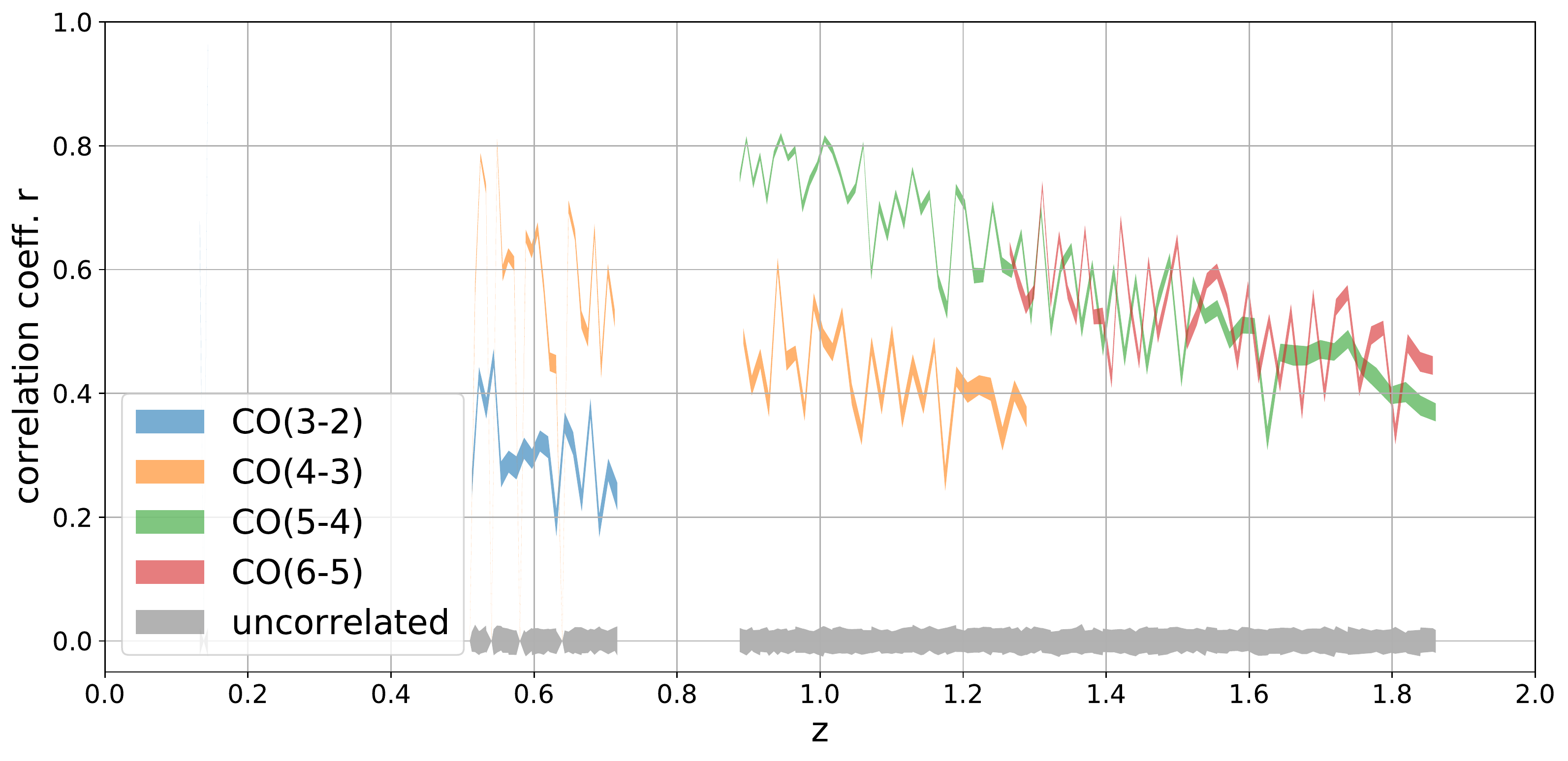}
\caption{\label{F:gen_r_txtn} Pearson correlation coefficient $r$ between the true line maps and the same maps adding the $\sigma_n=10^{-4}$ Jy sr$^{-1}$ noise on 2500 light cones. The bands are the rms of the value in 100 noise realizations with the sample line signal. The gray bands are the $r$ value shown with the uncorrelated white noise map for reference.}
\end{center}
\end{figure}

In the case of no interloper lines (i.e. only single line emission in the data), the best estimator of the single-line intensity map is the observed map (regardless of foregrounds). To compare our reconstruction performance with this limiting case, we calculate the correlation coefficient between the input single-line map with the same single-line map plus the instrument noise.

Fig.~\ref{F:gen_r_txtn} shows the results using $\sigma_n=10^4$ Jy sr$^{-1}$.  Comparing to Fig.~\ref{F:gen_r}, which has the same noise level, the single-line-plus-noise case has a lower correlation than the reconstruction in Fig.~\ref{F:gen_r}. Especially for the fainter lines (e.g., CO(3--2) at $z\sim 0.6$ and CO(4--3) at $z\sim1.1$), our reconstruction map has a much better correlation. This can be explained by the fact that in our algorithm, the sources are detected in the template space rather than in a single voxel. That is, if a source can be observed in two frequency channels, we extract the source by projecting the signals in these two channels to the template space, which is effectively combining the information from both channels. Consequently, we are able to achieve a better S/N on the fainter lines because of the greater sensitivity of their brighter counterpart.

\subsection{Improving Cross-correlation Uncertainty}
\begin{figure}[ht!]
\begin{center}
\includegraphics[width=\linewidth]{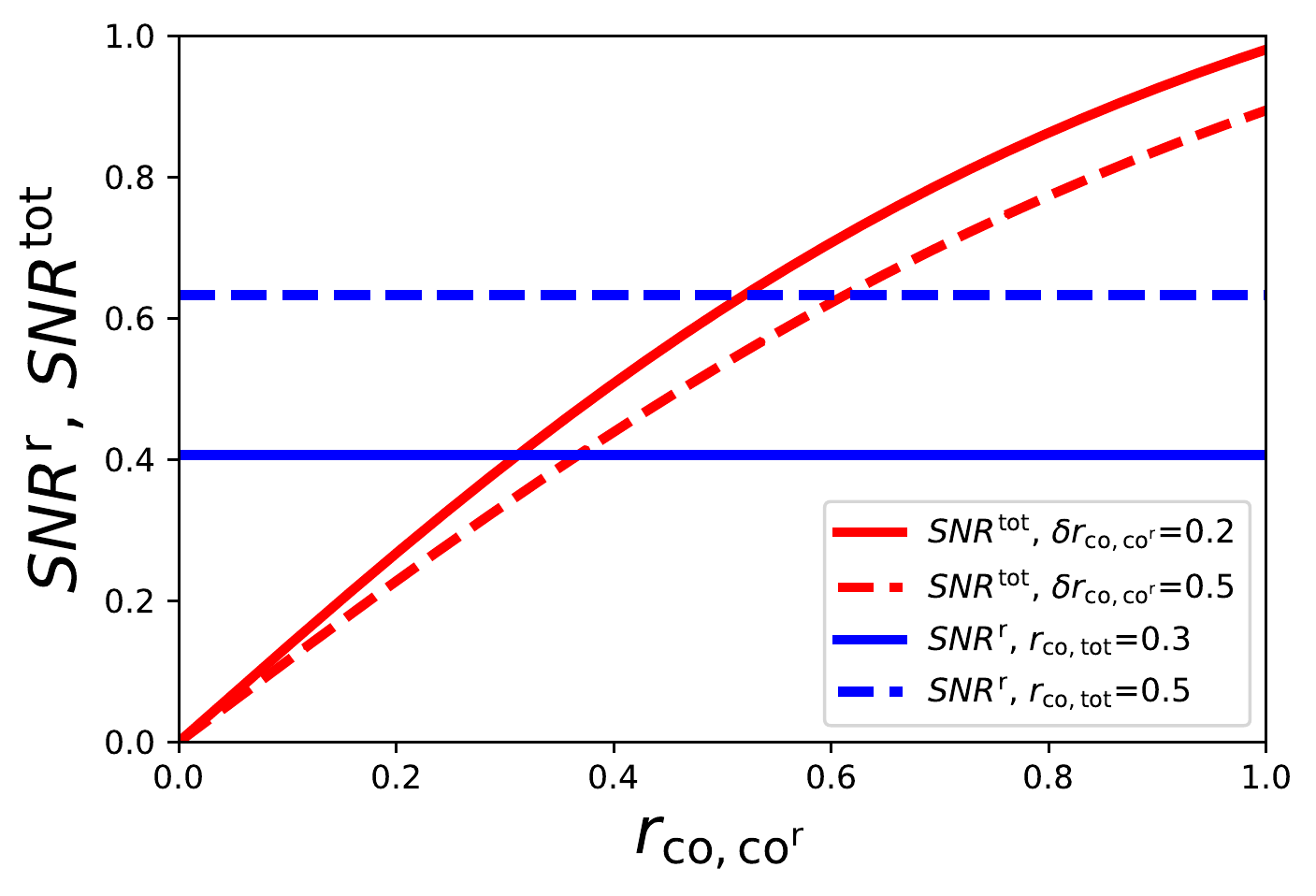}
\caption{\label{F:SNRxcorr} Comparing the S/N on the cross-power spectrum of one CO line and the external tracer, using the total observed map ($S/N^{\rm tot}$) and the reconstructed CO map ($S/N^{\rm r}$). We consider two different values of $\delta r_{\rm co,co^{r}}$ and $r_{\rm co,tot}$ that covers the range of realistic parameter values in our model. The value of $r_{\rm co,co^{r}}\sim 0.7-0.9$ according to the results in Sec~\ref{S:results}, and therefore using the reconstructed map instead of the total observed map in cross-correlation can reduce the uncertainty (i.e. $SNR^{\rm r} > SNR^{\rm tot}$).}
\end{center}
\end{figure}

\begin{figure}[ht!]
\begin{center}
\includegraphics[width=\linewidth]{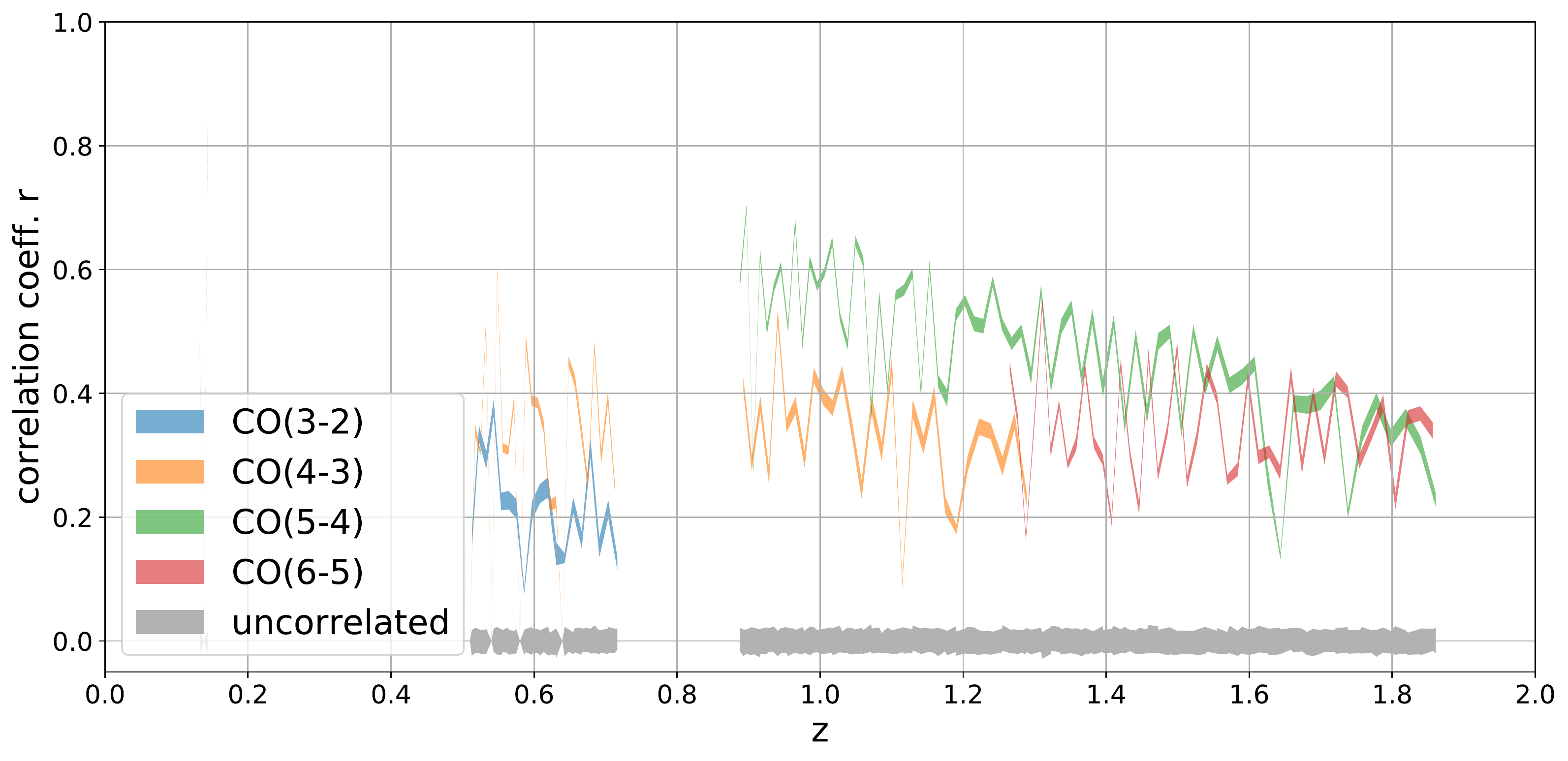}
\caption{\label{F:gen_r_all} Pearson correlation coefficient $r$ between the true input and the total observed maps on 2500 light cones with $\sigma_n=10^4$ Jy sr$^{-1}$. The bands are the rms of the value in 100 noise realizations with the sample line signal. The gray bands are the $r$ value shown with the uncorrelated white noise map for reference.}
\end{center}
\end{figure}

Another useful application of the reconstruction technique is for a more precise measurement of cross-correlation between LIM and other tracers. Cross-correlation analysis not only serves as a validation of a cosmological signal in LIM, since the cross-correlation is less susceptible to foreground contamination and other systematic effects, but also provides valuable astrophysical and cosmological information. 

While cross-correlating the total observed LIM map with an external tracer gives an unbiased estimator of cross spectrum between the target line and the external tracer, the presence of continuum foregrounds and interlopers increases the error of this measurement. Here we present a simple argument that for cross-correlation analysis, using a reconstructed map instead of the total observed map can effectively reduce the error bars on the cross-power spectrum. 

The cross-power spectrum errors between two fields $\delta P_{12}$ in a single mode are given by
\begin{equation}\label{E:xerr}
\delta P_{1,2}^2 = \frac{1}{2}\left ( P_{1,2}^2 + \delta P_1\delta P_2 \right ),
\end{equation}
where $P_{1,2}$ is the cross-power spectrum, and $\delta P_1$ and $\delta P_2$ are the errors on the auto power spectrum in the two fields. For a single $k$ mode, $\delta P_1=P_1$, $\delta P_2=P_2$, where $P_1$ and $P_2$ are the total power spectra (including signals and noise) measured in two fields. Eq.~\ref{E:xerr} can be expressed in terms of the cross-correlation coefficient: $r_{1,2}=P_{1,2}/\sqrt{P_1P_2}$,
\begin{equation}\label{E:xerr2}
\delta P_{1,2}^2 = \frac{1}{2}P_{1,2}^2\left (1+\frac{1}{r_{1,2}^2}\right ),
\end{equation}

Say we have an external galaxy sample that traces one of the target CO lines, then we can write the total observed LIM data as the combination of target CO line ($I_{\rm CO}$), other interloper lines ($I_{\rm interlopers}$), and the noise ($I_n$),
\begin{equation}
I_{\rm tot} = I_{\rm CO} +I_{\rm interlopers} +I_n.
\end{equation}
The expectation value of the cross spectrum between the observed total map and galaxy is the same as the cross spectrum with only the target CO, since the other components are not correlated with the large-scale structure at the same redshift, so $\left \langle P_{\rm g,tot} \right \rangle=\left \langle P_{\rm g,co} \right \rangle$. 

For simplicity, we assume the galaxies are perfectly correlated with the target CO line field on the scale of interest, so $r_{\rm g,co}=1$. This implies the galaxy field and CO field always have the same correlation $r$ with any given field $x$, $r_{\rm g,x}=r_{\rm co,x}$.

From Eq.~\ref{E:xerr2}, the error on the galaxy-CO cross spectrum measured by cross-correlating the galaxy field with the total observed LIM data is
\begin{equation}
\begin{split}
\delta P_{\rm g,co}^{\rm tot} &= P_{\rm g,tot}\sqrt{\frac{1}{2}\left(1+\frac{1}{r_{\rm g,tot}^2}\right)}\\
&= P_{\rm g,co}\sqrt{\frac{1}{2}\left(1+\frac{1}{r_{\rm co,tot}^2}\right)}
\end{split}
\end{equation}

On the other hand, if we cross-correlate the galaxies with the reconstructed CO map (${\rm CO^r}$), the error is
\begin{equation}
\delta P_{\rm g,co{^r}} = P_{\rm g,co{^r}}\sqrt{\frac{1}{2}\left(1+\frac{1}{r_{\rm g,co{^r}}^2}\right)}.
\end{equation}
However, $P_{\rm g,co{^r}}$ is a biased estimator of $P_{\rm g,co}$ because of the error in the reconstruction. If we assume that the reconstructed map roughly preserves the same power as the true map, $P_{\rm co^{r}}\approx P_{\rm co}$, then we can write\footnote{$P_{\rm g,co{^r}}=r_{\rm g,co{^r}}\sqrt{P_gP_{\rm co^{r}}}\approx r_{\rm g,co{^r}}\sqrt{P_gP_{\rm co}}=r_{\rm g,co{^r}}P_{\rm g-co}$, where the last equality uses the assumption $r_{\rm g,co}=1$.} $P_{\rm g,co{^r}}=r_{\rm g,co{^r}}P_{\rm g,co}=r_{\rm co,co{^r}}P_{\rm g,co}$. Therefore, we have to de-bias cross spectrum $P_{\rm g,co{^r}}$ by factor $1/r_{\rm co,co{^r}}$, $P_{\rm g,co}=P_{\rm g,co{^r}}/r_{\rm co,co{^r}}$. The value of $r_{\rm co,co{^r}}$ cannot be directly inferred from the data, so we have to estimate it by simulating the possible range of signals; this introduces an extra error term to the $\delta r_{\rm co,co{^r}}$ due to the uncertainty in $r_{\rm co,co{^r}}$,
\begin{equation}
\begin{split}
\delta P_{\rm g,co}^{\rm r} &= P_{\rm g,co}\sqrt{\left ( \frac{\delta P _{\rm g,co^{r}}}{P_{\rm g,co^{r}}} \right )^2 + \left ( \frac{\delta r _{\rm co,co^{r}}}{r_{\rm co,co^{r}}} \right )^2}\\
&=P_{\rm g,co}\sqrt{\frac{1}{2}\left ( 1+\frac{1}{r^2_{\rm co,co^{r}}} \right ) + \left ( \frac{\delta r _{\rm co,co^{r}}}{r_{\rm co,co^{r}}} \right )^2}.
\end{split}
\end{equation}

In summary, the S/N on the galaxy-CO cross spectrum $P_{\rm g,co}$ using the total observed map and the reconstructed map is
\begin{equation}
\begin{split}
SNR^{\rm tot}=\frac{P_{\rm g,co}}{\delta P_{\rm g,co}^{\rm tot}}&=\frac{1}{\sqrt{\frac{1}{2}\left(1+\frac{1}{r_{\rm co,tot}^2}\right)}},\\
SNR^{\rm r}=\frac{P_{\rm g,co}}{\delta P_{\rm g,co}^{\rm r}}&=\frac{1}{\sqrt{\frac{1}{2}\left ( 1+\frac{1}{r^2_{\rm co,co^{r}}} \right ) + \left ( \frac{\delta r _{\rm co,co^{r}}}{r_{\rm co,co^{r}}} \right )^2}}.
\end{split}
\end{equation}

Note that the S/N in both cases converges to unity when $r$ is unity and the de-bias error is zero, which is the limit of sample variance\footnote{The power spectrum cross-correlation coefficient can be derived from the Pearson correlation (Eq.~\ref{E:pearson}) with a weighting on pixels. Since our pixels are generated and reconstructed independently of each other, the Pearson correlation coefficient here is an unbiased estimator of the power spectrum correlation coefficient. Thus here we will use the value of Pearson correlation we derived for the power spectrum correlation coefficient.}. 

Fig.~\ref{F:SNRxcorr} shows the $S/N^{\rm tot}$ and $S/N^{\rm r}$ with two different values of $\delta r_{\rm co,co^{r}}$ and $r_{\rm co,tot}$ as a function of $r_{\rm co,co^{r}}$. According to the calculations in Sec.~\ref{S:model_uncertainty}, the value of $r_{\rm co,co^{r}}$ ranges from $\sim 0.7$ to $\sim 0.9$ (except for the most noisy case, $\sigma_n=5\times 10^4$ Jy sr$^{-1}$). To estimate the realistic $r_{\rm co,tot}$ value, we calculate the correlation of the input CO line maps and the observed map with $\sigma_n=10^4$ Jy sr$^{-1}$, the same noise level as in Fig.~\ref{F:gen_r}. The results are shown in Fig.~\ref{F:gen_r_all}. We find that $r_{\rm co,tot}$ are around 0.2-0.5 in this case. With this range of parameters, Fig.~\ref{F:SNRxcorr} indicates that $S/N^{\rm r}$ is better than $S/N^{\rm tot}$, which means that in our model, using the reconstructed map instead of the observed map in cross-correlation can reduce the uncertainty. 

\subsection{Estimating the complete interloper population}
Here we discuss a potential extension of the technique, which is capable of estimating the {\it total} power of an interloper population.  The method uses an incompletely reconstructed interloper sample and an incomplete, external  tracer of the interloper density field. Once the interloper contribution is fully quantified, the high-redshift signal of interest in an LIM dataset can be estimated without bias. As an example, we write the observed intensity of a [\ion{C}{ii}] LIM dataset as 
\begin{equation}
    I^{\rm obs} = I_{\rm CO} + I_{\rm C\,II} + \delta_n^{\rm obs},
\end{equation}
where $I_{\rm CO}$ is the total  CO interloper intensity, $I_{\rm C\,II}$ is the \ion{C}{ii} signal, and $\delta_n^{\rm obs}$ is the instrumental noise. For simplicity, here we only consider the contribution of one CO rotational line as the foreground.

Given a reconstruction threshold, we reconstruct the bright CO emissions using our technique as:
\begin{equation}
    I^{\rm rec}_{\rm CO} = \alpha I_{\rm CO_b} + \delta_n^{\rm rec}, 
\end{equation}
where $I^{\rm rec}_{\rm CO}$ is the reconstructed CO intensity and $I_{\rm CO_b}$ is the intensity of bright CO sources in the reconstruction, which is a subset of the total CO population. There are two sources of error in the reconstruction: a multiplicative term $\alpha$ proportional to the bright CO intensity, where $\alpha$ can be greater or smaller than unity, and an additive term $\delta_n^{\rm rec}$ describing random (or misidentified) fluctuations about the true CO intensity, which is uncorrelated with the CO field. 

The auto power spectrum of the reconstructed CO map is
\begin{equation}
\begin{split}
    \left \langle  I_{\rm CO}^{\rm rec}I_{\rm CO}^{\rm rec} \right \rangle =& \bar{\alpha}^2 \left \langle  I_{\rm CO_b}I_{\rm CO_b} \right \rangle  + \left \langle \delta_n^{\rm rec}\delta_n^{\rm rec} \right \rangle \\
    =& \bar{\alpha}^2  \left \langle  I_{\rm CO_b} \right \rangle^2 b_{\rm CO_b}^2 P(k) + N^{\rm rec},
\end{split}
\end{equation}
where, in the linear regime that we consider here, $b_{\rm CO_b}$ is the cosmological clustering bias of the bright CO sources, $P(k)$ is the matter density field, and $N^{\rm rec}$ is the auto power spectrum of the $\delta_n^{\rm rec}$ term as a noise bias.

Using an external galaxy sample $g$ in the same redshift range as the CO interlopers, we cross-correlate $g$ with the observed and reconstructed maps, respectively, and consider only linear clustering scales:
\begin{equation}
    \left \langle g I_{\rm obs} \right \rangle  = r \left \langle  I_{\rm CO} \right \rangle b_{\rm CO} b_{\rm g} P(k),
\end{equation}
\begin{equation}
    \left \langle g I_{\rm CO}^{\rm rec} \right \rangle = \left \langle g \alpha I_{\rm CO_b} \right \rangle = \bar{\alpha} r_b \left \langle  I_{\rm CO_b} \right \rangle b_{\rm CO_b} b_{\rm g} P(k), 
\end{equation}
where $\langle \alpha \rangle = \bar{\alpha}$, $b_g$ is the bias of the galaxy tracer, $b_{\rm CO_b}$ is the bias of the bright CO population, and $r$ ($r_b$) is the astrophysical stochastic cross-correlation parameter between the galaxy and CO (CO$_b$) populations.  

Finally, we construct an estimator of the full CO power spectrum as
\begin{equation}\label{E:Pco_est}
\begin{split}
   &\left \langle  \widehat{I_{\rm CO} I_{\rm CO}} \right \rangle = \left \langle  I_{\rm CO}^{\rm rec}I_{\rm CO}^{\rm rec} \right \rangle\left (  \frac{\left \langle g I_{\rm CO}^{\rm obs} \right \rangle}{\left \langle g I_{\rm CO}^{\rm rec} \right \rangle} \right )^2 \\
   &= \left (\bar{\alpha}^2 \left \langle  I_{\rm CO_b} \right \rangle^2 b_{\rm CO_b}^2 P(k) + N^{\rm rec}  \right ) \left ( \frac{r\left \langle I_{\rm CO}\right \rangle b_{CO}}{\bar{\alpha} r_b\left \langle I_{\rm CO_b}\right \rangle b_{CO_b} } \right )^2 \\
   &= \left ( \left \langle  I_{\rm CO_b} \right \rangle^2 b_{\rm CO_b}^2 P(k) + \frac{N^{\rm rec}}{\bar{\alpha}^2}  \right ) \left ( \frac{r\left \langle I_{\rm CO}\right \rangle b_{CO}}{ r_b\left \langle I_{\rm CO_b}\right \rangle b_{CO_b} } \right )^2 \\
\end{split}
\end{equation}

% On large scales, $r = r_b = 1$, as all tracers follow the dark matter distribution. If we carefully pick a reconstruction threshold such that the reconstruction of a subset of the CO population (bright CO sources) is nearly noiseless, $N^{\rm rec}/\bar{\alpha}^2 \rightarrow 0$, this estimator becomes an  unbiased estimator of the full CO power spectrum on large scales: 
If we can pick a reconstruction threshold such that the reconstruction of a subset of the CO population (bright CO sources) includes the majority of CO emitters, then $r \approx r_b$. Furthermore, if the reconstruction noise is negligible, $N^{\rm rec}/\bar{\alpha}^2 \rightarrow 0$, then this estimator becomes an  unbiased estimator of the full CO power spectrum on large scales:
\begin{equation}
    \left \langle  \widehat{I_{\rm CO} I_{\rm CO}} \right \rangle =
   \left \langle  I_{\rm CO}^{\rm rec}I_{\rm CO}^{\rm rec} \right \rangle\left (  \frac{\left \langle g I_{\rm CO}^{\rm obs} \right \rangle}{\left \langle g I_{\rm CO}^{\rm rec} \right \rangle} \right )^2 = \left \langle  I_{\rm CO} \right \rangle^2 b_{\rm CO}^2 P(k).
\end{equation}
In principle, this argument holds regardless of the luminosity limit of the CO population used for the reconstruction {\it and} regardless of the magnitude limit of the galaxy sample used for the cross-correlation estimate, as long as the galaxy and CO samples overlap spatially. This is potentially a powerful approach to access the entire interloper CO population without the need to identify  the faint, undetected source contributions. Subtracting this CO estimate then provides an unbiased estimate of the high-redshift \ion{C}{ii} power spectrum in the same LIM dataset, which is highly desirable. 

With an external galaxy catalog along, one can also estimate the total CO power on  large scales where $r\approx 1$, with the estimator: $\left \langle g I_{\rm CO}^{\rm obs} \right \rangle^2 / \left \langle g g \right \rangle$. However, the advantage of the  Eq.~\ref{E:Pco_est} estimator is that it only requires $r \approx r_b$, which can be valid on smaller scales where $r < 1$, as long as the reconstruction has sufficient quality in terms of the noise level purity
and completeness. Therefore, with the reconstructed map, one can potentially extract the total CO power to smaller scales.

We note that a high reconstruction threshold $m$, is required in order to achieve a (nearly) noiseless reconstruction ($N^{\rm rec}/\bar{\alpha}^2 \rightarrow 0$). On the other hand, $r \approx r_b$ becomes invalid if the threshold is too high such that majority of the fainter sources is missed in the reconstruction. In addition, a high threshold tends to boost the shot noise in the reconstructed  power spectrum, as well as uncertainties in the CO power spectrum estimator (Eq.~\ref{E:Pco_est}). There is clearly a trade-off between the fidelity of the reconstructed signals and the uncertainty to estimate the desired signals. A detailed simulation is necessary to determine the optimal threshold to minimize the effect of bias from $N_{\rm rec}$ and the variance from the reconstruction shot noise, and evaluate the performance of the CO power spectrum estimator. We leave this investigation to future work.

\subsection{Comparing Foreground Cleaning Capability with Masking}
\begin{figure*}[ht!]
\includegraphics[width=\linewidth]{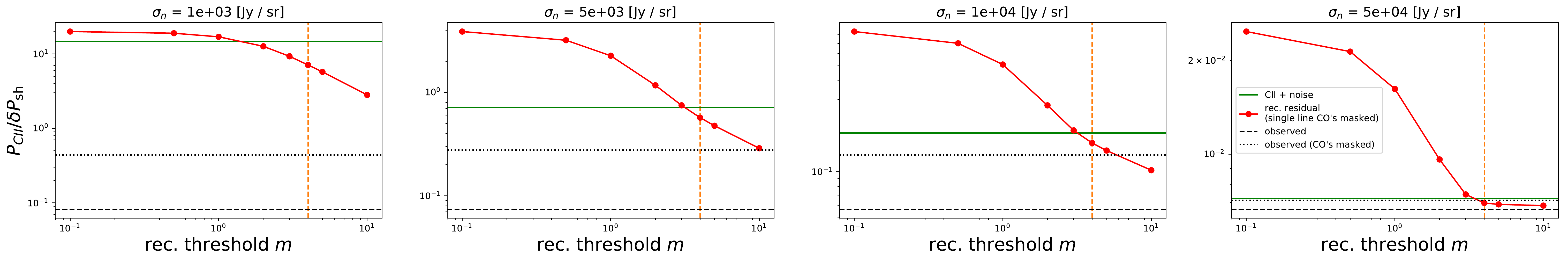}
\caption{\label{F:res_mask} S/N on [\ion{C}{ii}] shot-noise power spectrum after CO line foreground removal with masking (black dotted) and sparse reconstruction (red) with different instrument noise level $\sigma_n$ and reconstruction threshold $m$. The orange dashed line marks the $m=4$ threshold for reference. We also show the S/N before cleaning (black dashed), and the limiting case of the no CO signals (green) for comparison.}
\end{figure*}

Our algorithm can serve as a foreground mitigation method for \ion{C}{ii} LIM measurement by identifying the bright CO foreground signals. In this section, we quantify this "foreground cleaning" performance using our algorithm and compare it with the masking method, where a ``cleaned map'' is obtained by masking out voxels that contain bright CO sources identified with an external source catalog. 

In the following, we compare two cases: (1) in the masking case, the ``cleaned map'' is the observed map (including all of the lines and instrument noise) masked using external CO catalogs; and (2) the ``cleaned map'' derived from our algorithm is the observed map subtracting the CO reconstructed map. 

The external catalog considered here is the same as the one described in Sec.~\ref{S:ext_cats}: a flux cut on CO(5--4) at the level of $150\, L_\odot$ Mpc$^{-2}$ ($6\times 10^{-17}$ W m$^{-2}$), which gives $\sim 1.2$ galaxies per light cone ($0.43^2$ arcmin$^2$ pixel solid angle) and corresponds to a $m_{\rm AB}\sim 21.8$ threshold in the optical band. Note that this masking threshold is comparable to the ``case A'' masking in \citet{2018ApJ...856..107S}, although here we consider a simpler model that ignores the scatter in the line luminosity model. For this masking scenario, the ``cleaned map'' is the observed map minus any voxels that contain the sources in the external catalog.

In our algorithm, the CO sources and their spectra are reconstructed and removed from the data iteratively, and the residual can be regarded as a ``cleaned map'' that is free of bright CO sources. However, we are only capable of cleaning the multi-line redshift bins in our algorithm. The signal identified in the single-line redshift bins in our algorithm is the combination of the remaining CO and \ion{C}{ii} signals. If we remove all of the reconstructed single-line signals, we will over-subtract \ion{C}{ii} in the cleaned map. Therefore, for the single-line redshift bins ($0.15\leq z\leq0.51$; $0.72\leq z\leq0.89$; $ z\geq1.87$), we clean the data by masking  voxels that contain external catalog sources.

As a figure of merit, we calculate the S/N of the respective \ion{C}{ii} shot-noise power spectra (since there is no clustering signal in our mock light cones) while including the residual CO as part of the noise contribution. 

The error on the shot-noise power spectrum $\delta P_{\rm sh}$ in a map is
\begin{equation}
\delta P_{\rm sh} = \frac{P_{\rm tot}}{\sqrt{N_{\rm mode}}}=\frac{V_{\rm vox}\sigma^2_{\rm tot}}{\sqrt{N_{\rm mode}}},
\end{equation}
where $P_{\rm tot}$ is the total power spectrum of the cleaned map on the shot-noise scales, which is proportional to the total voxel variance in the map, $\sigma^2_{\rm tot}$. $N_{\rm mode}$ is the number of k-space modes used to measure the shot noise. $N_{\rm mode}$ is usually of the order of the total number of voxels, so we choose $N_{\rm mode}=6000$, similar to the number of voxels in TIME. The shot-noise power of the \ion{C}{ii} signal is given by
\begin{equation}
 P_{\rm CII}=V_{\rm vox}\sigma^2_{\rm CII},
\end{equation}
where $\sigma^2_{\rm CII}$ is the voxel variance of the \ion{C}{ii} signal map. The S/N of the \ion{C}{ii} shot-noise power spectrum is then:
\begin{equation}\label{E:CII_SNR}
 \frac{P_{\rm CII}}{\delta P_{\rm sh}}=\frac{\sigma^2_{\rm CII}}{\sigma^2_{\rm tot}}\sqrt{N_{\rm mode}}.
\end{equation}

We test the foreground cleaning performance with the same set of 2500 mock light cones described in Sec.~\ref{S:results}, and calculate the \ion{C}{ii} shot-noise S/N using Eq.~\ref{E:CII_SNR}. The $\sigma^2_{\rm CII}$ is the variance of the input \ion{C}{ii} map, and $\sigma^2_{\rm tot}$ is the variance of the cleaned map. For simplicity, $\sigma^2_{\rm CII}$ and $\sigma^2_{\rm tot}$ are the variance from all of the frequency channels.

In Fig.~\ref{F:res_mask}, the black dotted lines and the red lines show the \ion{C}{ii} shot-noise S/N with the cleaned map obtained from masking and from our algorithm, respectively. For reference, the black dashed lines are the \ion{C}{ii} shot-noise S/N of the observed map before cleaning, and thus the map includes the contribution from all of the lines and the instrument noise.

According to Fig.~\ref{F:gen_r_bands}, the optimal threshold that gives the maximum $r$ value is $m\sim4$. In Fig.~\ref{F:res_mask}, $m=4$ is marked with an orange dashed line, and we see that in the highest-noise case for this threshold, masking (black dotted line) using an external (deep) catalog  performs slightly better than reconstruction (red line) because it is difficult for the MP algorithm to extract the signals from noisy data directly. For realistic noise levels (between $\sigma_n=5\times10^3$ and $10^4$  Jy sr$^{-1}$), the reconstruction outperforms masking.

We also compare this result with the limiting case where there are only \ion{C}{ii} and instrumental noise in the data (green line), i.e. $\sigma^2_{\rm tot} = \sigma^2_{\rm CII} + \sigma^2_n$. For a  small reconstruction threshold $m$, the reconstructed S/N is better than this limit, which indicates that the reconstruction over-fits and misidentifies noise fluctuation as a signal and removes them from the cleaned data. We see that for the two realistic noise levels at $m\sim4$, the reconstructed S/N is lower than this limit, and thus indicates that overfitting is not an issue at this threshold. Also note that in the highest-noise case, the masking SNR is close to the noise-plus-\ion{C}{ii} limit (green line), for the following reason: since the external catalog goes much deeper than the noise level, noise fluctuation dominates over the line signals after masking, and thus variance in the masked map is close to the noise variance.

Finally, we point out that the reconstructed \ion{C}{ii} shot-noise S/N (red line) converges to a constant instead of increasing with smaller threshold $m$ values. This is because of the fact that in the low threshold limit, the reconstruction residual in the cleaned map is subdominant compared to the (masked) single-line redshift bin signals being added back to the residual, and thus the cleaned map S/N does not depend on the threshold value. 

To sum up, for a realistic noise level, our reconstruction performs better than masking in terms of foreground cleaning capability, given our signal model and the external catalog considered in this work. We note that this conclusion depends on the line luminosity function model and the depth of the external catalog for masking, and we leave a more detailed analysis to future work.

\subsection{Technique Extensions}
In this section, we outline some directions for extending the current framework to further improve upon the line reconstruction in future work. 

\subsubsection{Template Generalization}
Currently we use a single SLED template for all redshift bins in the reconstruction; this can be easily generalized to incorporate multiple spectral templates to account for the redshift evolution and SLED variation of the signals. In addition, the  extension can help differentiate the emission from different types of galaxies that have different SLEDs. For instance, if we have two different SLED models for early- and late-type galaxies, respectively, we can incorporate them by having two columns in $\mathbf{A}$ for every redshift bin such that the reconstruction can infer not only the redshift and luminosity of the sources, but also their galaxy type from the SLED templates. 

\subsubsection{Alternative Sparse Approximation Algorithm}
The MP algorithm adopted in this work optimizes the $\ell_0$ norm in Eq.~\ref{E:sparse_eqn}, which is the direct sum of the nonzero elements in $\widetilde{\mathbf{N}}$. We can improve the algorithm by including prior information on the expected value of each element in $\widetilde{\mathbf{N}}$, which depends on the voxel size and source luminosity function. For instance, instead of using the MP algorithm, one can obtain the sparse solution by solving the following  $\ell_1$-norm regularization equation:
\begin{equation}\label{E:L1_reg}
\underset{\widetilde{\mathbf{N}} }{\mathrm{argmin}}\left \| \mathbf{I} - \mathbf{A}\widetilde{\mathbf{N}} \right \|^2_2+\lambda\left \| \mathbf{w} \cdot\widetilde{\mathbf{N}} \right \|_1,
\end{equation}
where the parameter $\lambda$ determines the regularization strength for preventing overfitting, which has a similar effect as the stopping criteria in the MP algorithm; the prior information on the number density of the sources in each redshift bin can be encoded in the weight vector $\mathbf{w}$ in this expression.

\subsubsection{Clustering Information}
In this work, we only perform the pixel-by-pixel line de-confusion using the information in the spectral correlation due to the multiple lines emitted from the same source to reconstruct the signals. The clustering information of the galaxies, which is neglected in this work, could provide additional information on the emission field. We can generalize this framework by incorporating the clustering information from the known galaxy two-point correlation, and perform the reconstruction on an ensemble of pixels to simultaneously fit for the spectral correlation and clustering. For example, if we have a theoretical model for the line-of-sight two-point correlation function of $\mathbf{N}(z)$, $\xi_{\rm th}(\mathbf{N}(z))$, we can add another $\ell_2$-norm regularization term to Eq.~\ref{E:L1_reg},
\begin{equation}
    \lambda_{\rm clus}\left \| \xi(\widetilde{\mathbf{N}}(z)) - \xi_{\rm th}(\mathbf{N}(z))  \right \|_2^2.
\end{equation}
This will enforce the algorithm to give higher priority to solutions close to the theoretical correlation function. Similarly, with an external catalog that traces the same large-scale structure, one can also constrain the algorithm with cross-correlation:
\begin{equation}
    \lambda_{\rm clus}^{x}\left \| \xi^x(\widetilde{\mathbf{N}}(z), \delta_{\rm ext}) - \xi^x_{\rm th}(\mathbf{N}(z), \delta_{\rm ext})  \right \|_2^2,
\end{equation}
where $\delta_{\rm ext}$ is the density field of the external tracer, e.g. galaxy samples, and $\xi^x$ is the cross-correlation between $\mathbf{N}(z)$ and $\delta_{\rm ext}$. We leave further investigation that makes use of clustering information to future work.

\section{Conclusion}\label{S:conclusion}
We develop a spectral line de-confusion technique for LIM experiments, where different spectral lines emitted by sources at different redshifts can be observed in the same frequency channel and then confused. Unlike most of the previously proposed methods that decompose the line signals in the power spectrum space, we perform a phase-space de-confusion that reconstructs  the individual line intensity maps, if multiple spectral lines of a redshifted source population are observable. The reconstructed line intensity maps are direct data products of an LIM experiment, and can be used to trace the underlying density field for various science applications.

Our method is based on the information that multiple spectral lines emitted by redshifted sources are mapped onto distinct observed frequencies, which give deterministic features in the observed spectrum that can be fitted by a template. With a set of spectral template models and assuming the sparse approximation, we fit the LIM data iteratively with the MP algorithm.

As an example, we consider an LIM survey with similar survey parameters as the ongoing EoR [\ion{C}{ii}] experiments, TIME, and CONCERTO. The intervening CO line intensity maps at $0.5\lesssim z\lesssim1.5$ can be extracted with our technique, since multiple CO rotational transitions are observable. We demonstrate that with the assumed signal model and realistic noise level, our reconstructed CO maps reach $\sim 80\%$ spatial correlation with the true maps. In addition, in our assumed signal model and realistic survey setup, the VID of individual lines can be correctly extracted with a high S/N ratio down to the $\ell_*$-scale. The CO luminosity function derived from the VID measurement can provide information on galaxy formation and evolution as traced by the CO distribution across cosmic time. The reconstruction performance is robust against a realistic level of line ratio uncertainties and continuum foreground mitigation process.

In addition to probing  the large-scale luminosity and density fields, the reconstructed line intensity maps can also be used for a variety of applications. As a demonstration, we show that using the reconstructed map instead of the original LIM dataset can effectively reduce uncertainties in cross-correlation measurements, and improve the performance of interloper masking to reveal the high-redshift line emissions. Furthermore, given that the reconstructed intensity map, even if incomplete, traces the matter density on large scales, we construct an estimator capable of estimating the total interloper power. The estimator invokes the cross-correlation of the reconstructed map with an external density tracer such as galaxies in the linear clustering regime. This approach has the potential to fully specify the interloper and high-redshift source populations and warrants future investigation.

While we mainly discuss the application for an EoR [\ion{C}{ii}] LIM experiment in this paper, this technique is not restricted to this setup. We demonstrate that our technique can successfully extract redshifted optical line signals from a SPHEREx-like experiment in the near-infrared. The technique is a general framework that can be readily applied to mitigate line de-confusion problems in LIM experiments and enhance the science returns.

\acknowledgments
We are grateful for the helpful discussions with Chun-Lin Liu, Patrick Breysse, Matthieu B{\'e}thermin, Emmanuel Schaan, Matt Orr, the TIME collaboration, the Caltech ObsCos group, and the participants of the conference ``Lines in the Large Scale Structure.'' We would like to thank the anonymous referee for
valuable comments that improved the manuscript. Part of the research described in this paper was carried out at the Jet Propulsion Laboratory, California Institute of Technology, under a contract with the National Aeronautics and Space Administration.

\appendix
\section{The Matching Pursuit (MP) Algorithm}\label{A:MPA}
\begin{figure*}[htbp!]
\begin{center}
\includegraphics[width=\linewidth]{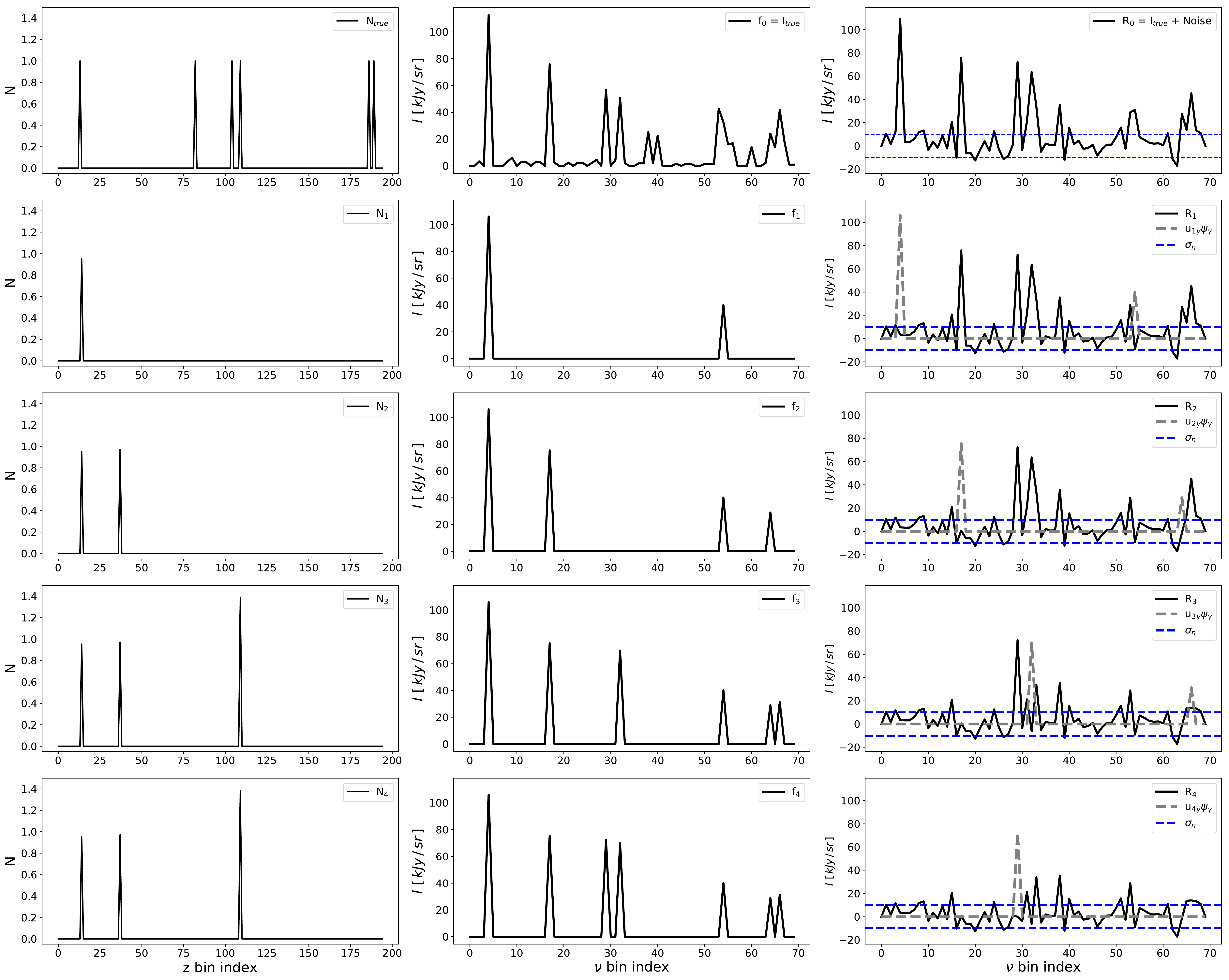}
\caption{\label{F:MP_steps} Illustration of how the MP algorithm solves for the source vector $\mathbf{N}$. See the text for a detailed description.}
\end{center}
\end{figure*}

In this section, we describe the detailed steps in the matching pursuit (MP) algorithm. The MP algorithm iteratively selects an atom in the dictionary to project out part of the signals in the data, and keep track of the current solution of the signal $\mathbf{f}$ and residual $\mathbf{R}$ for the next step until the solution meets the stopping criteria. In Eq.~\ref{E:sparse_eqn}, for a given signal $\mathbf{I}$ and matrix $\mathbf{A}$, we define a set of vectors $\{\mathbf{\psi}_i\}$ to be the column vectors in $\mathbf{A}$ (i.e. the atoms in the dictionary). The MP algorithm works as follows:

\begin{enumerate}
\item Initialize at step $t=0$: $\mathbf{f}_0 = \mathbf{0}$, $\mathbf{R}_0 = \mathbf{I}$, $\widetilde{\mathbf{N}}_0 = \mathbf{0}$.
\item Compute the inner product of $\mathbf{R}_0$ and $\psi_i$'s:
\begin{equation}
\left \{ u_{0i} \right \} = \left \{ \left \langle \mathbf{R}_0, \psi_i \right \rangle \right \}.
\end{equation}
\item Select the element $\gamma$ to be updated by 
\begin{equation}
\gamma = \underset{i}{\mathrm{argmax}}\left \{ u_{0i} \right \}.
\end{equation}
\item If $u_{0\gamma}$ meets the stopping criteria, end the process and return $\mathbf{N} = \mathbf{0}$. Else, proceed to step 5.
\item Update the current $\mathbf{f}_t$, $\mathbf{R}_t$, and record the amplitude of the new solution in $\gamma$-th element of vector $\widetilde{\mathbf{N}}_t$:
\begin{align}
\mathbf{f}_{t+1} &= \mathbf{f}_t + u_{t\gamma}\mathbf\,{\psi}_{\gamma},\\
\mathbf{R}_{t+1} &= \mathbf{R}_t - u_{t\gamma}\mathbf\,{\psi}_{\gamma},\\
\widetilde{\mathbf{N}}_{t+1}(\gamma) &= \widetilde{\mathbf{N}}_t(\gamma) +u_{t\gamma}
\end{align}
\item Compute the inner product of $\mathbf{R}_{t+1}$ and $\psi_i$'s:
\begin{equation}
\left \{ u_{(t+1)i} \right \} = \left \{ \left \langle \mathbf{R}_{t+1}, \psi_i \right \rangle \right \}.
\end{equation}
\item Select the element $\gamma$ to be updated by 
\begin{equation}
\gamma = \underset{i}{\mathrm{argmax}}\left \{ u_{(t+1)i} \right \}.
\end{equation}
\item If $u_{(t+1)\gamma}$ meets the stopping criteria, go back to step 5 for the next iteration. Else, proceed to step 9.
\item Return the final solution $\mathbf{N} = \widetilde{\mathbf{N}}_{t+1} \,/ \,\mathbf{I}^{\rm norm}$.
\end{enumerate}

As described in Sec.~\ref{S:methods_MP}, the stopping criteria is set by comparing $u_{t\gamma}$ with the noise $\sigma_n$. This follows the fact that var$ (u_{t\gamma}) = \sigma_n^2$ (see Appendix~\ref{A:proof_u_var} for the proof), so if we set $u_{(t+1)\gamma} < m\,\sigma_n$ in step 8, this is effectively setting an ``m-$\sigma$'' detection threshold (e.g., $m=5$ for a 5$\sigma$ detection).

Fig.~\ref{F:MP_steps} illustrates the steps of the MP algorithm solving $\mathbf{N}$ of an example light cone. In this example, we set $\sigma_n=10$ kJy, and the detection threshold $m=5$. In this light cone, there are six $\ell_*$ sources in the multi-line redshift bins at $z =$ [0.54, 1.06, 1.20, 1.24, 1.79, 1.82], shown in the top left panel. Since $\mathbf{N}$ is the effective number of $\ell_*$ sources per redshift bin, the amplitudes in the six corresponding redshift bins are equal to unity. The top middle panel is the total line signal in this light cone $\mathbf{I}_{\rm true}$, including the emission from those six sources as well as that from other sources in the single-line redshifts. The top right panel shows the observed data after adding noise to the signal $\mathbf{I}_{\rm true}$, which is also the $\mathbf{R}_0$ vector in the first step of the MP algorithm. The blue dashed lines mark the noise level $\pm \sigma_n = 10$ kJy for reference. 

In the first iteration, the MP algorithm selects the 14th z-bin index ($z = 0.54$) with amplitude $u_{1\gamma}\sim 1$, so the $z=0.54$ source is successfully extracted in this iteration. The gray dashed spectrum in the left panel of the second row is the template signal extracted in this step ($u_{1\gamma}\psi_\gamma$). 
The updated values of $\mathbf{N}_1$, $\mathbf{f}_1$, and $\mathbf{R}_1$ from step 5 are shown in the second row of Fig.~\ref{F:MP_steps}.

Then we proceed to the second iteration. The MP algorithm selects the 37th column ($z = 0.65$). The gray dashed spectrum in the left panel of the third row is the template signal fitted in this step ($u_{2\gamma}\psi_\gamma$). However, there is no $z=0.65$ source in the input, which means the MP algorithm misidentifies the emission from noise or other lines as the signal. The third row of Fig.~\ref{F:MP_steps} shows the updated values of $\mathbf{N}_2$, $\mathbf{f}_2$, and $\mathbf{R}_2$ from the second iteration. 

In the third iteration, the 109th ($z = 1.24$) column in the dictionary is selected. 
The fourth row of Fig.~\ref{F:MP_steps} shows the updated values of $\mathbf{N}_3$, $\mathbf{f}_3$, and $\mathbf{R}_3$ from this iteration. Note that this time the MP algorithm picks up a correct redshift, while it overestimates the amplitude by $\sim 40 \%$.

In the fourth iteration, the algorithm selects the 224th column from the dictionary. The 224th column is not in the multi-line redshift bins (first 195 columns), and thus the $\mathbf{N}_4$ in the bottom row of Fig.~\ref{F:MP_steps} remains unchanged, whereas $\mathbf{f}_4$ and $\mathbf{R}_4$ are updated with a single-peak signal.

In the fifth iteration, the stopping criteria in step 8 is met ($u_{5\gamma} < 5\sigma_n$), so the reconstruction terminates and returns the last row of Fig.~\ref{F:MP_steps} as the reconstruction results for this light cone.

In summary, in this example, two of the six multi-line redshift sources have been reconstructed, in addition to one misidentified source. Comparing the final reconstructed light cone signal (bottom middle panel of Fig.~\ref{F:MP_steps}) to the true input light cone (top middle panel of Fig.~\ref{F:MP_steps}), we can see that the MP reconstruction captures the strong peaks in the data, and the remaining signals are close to the noise level.

\def\varu{{\rm var}(u_{t\gamma}) = \sigma_n^2}
\section{proving $\varu$}\label{A:proof_u_var}
In Sec.~\ref{A:MPA}, the  residual of step $t$ $\mathbf{R_t}$ can be expressed in the linear combination of the dictionary atoms and noise:
\begin{equation}
\mathbf{R}_t = \sum_i c_i\,\psi_i + \mathbf{n},
\end{equation}
where $c_i$'s represents the constant coefficient. Then we derive
\begin{equation}
\begin{split}
u_{t\gamma} &\equiv \langle\ \mathbf{R}_t, \psi_\gamma\rangle\\
&= \sum_i c_i \langle\ \psi_i, \psi_\gamma\rangle + \langle\ \mathbf{n}, \psi_\gamma\rangle\\
&={\rm const} + \sum_j\,n_j \,\psi_{\gamma j}.
\end{split}
\end{equation}
The first term is not depend on the noise, so it is a constant term that is not contributing to the variance. Also note that $ \langle\ \psi_i, \psi_\gamma\rangle\neq \delta_{i\gamma}$ since the dictionary $\{\psi_i\}$ are normalized but not orthogonal. With this expression, we can calculate the variance:
\begin{equation}
\begin{split}
\left \langle u_{t\gamma} \right \rangle &={\rm const} + \sum_j\,\left \langle n_j \right \rangle \,\psi_{\gamma j} = {\rm const}\\
\left \langle u_{t\gamma}^2 \right \rangle &={\rm const}^2 + \sum_j\,\left \langle n_j^2 \right \rangle \,\psi_{\gamma j}^2\\
&={\rm const}^2 + \sigma_n^2\, \sum_j\,\psi_{\gamma j}^2\\
&={\rm const}^2 + \sigma_n^2.
\end{split}
\end{equation}
Therefore, we get
\begin{equation}
{\rm var}(u_{t\gamma}) = \left \langle u_{t\gamma}^2 \right \rangle - \left \langle u_{t\gamma} \right \rangle^2 = \sigma_n^2
\end{equation}

\section{SPHEREx Line Signal Model}\label{A:SPHEREx}
In this section, we describe the line signal model in the SPHEREx wavelengths. We model five lines from $z=0$ to 10 in SPHEREx band: Ly$\alpha$ (121.6 nm), H$\alpha$ (656.3 nm), H$\beta$ (486.1 nm), [\ion{O}{ii}] (372.7 nm), and [\ion{O}{iii}] (500.7 nm).

Since the optical lines are associated with the star formation activities, we model the signal with the following steps: we start with the halo mass function, and use the star formation rate (SFR)--halo mass ($M$) relation, and the SFR--line luminosity relation to paint the spectral line signals to each halo.

We use the publicly available halo mass function calculator \textit{HMFcalc} \citep{2013A&C.....3...23M}\footnote{\url{http://hmf.icrar.org/}} to obtain the halo mass function based on the \citet{2001MNRAS.323....1S} model. For the SFR--M relation, we use the model from \citet{2013ApJ...770...57B}, in which the SFR--M relation is derived based on several observational constraints. \footnote{The $SFR(M, z)$ is downloaded from the author's webpage (\url{https://www.peterbehroozi.com/data.html}). 
The model is only available at $0<z<8$, so we use $z$ = 8 model for $z>8$.}

For the SFR--line luminosity relation, we assume a linear relation for all of the lines. For Ly$\alpha$, we use the prescription provided by \citet{2017MNRAS.464.1948F} with their fiducial values: $\gamma_{\rm{Ly}\alpha} = 1, f^{UV}_{\rm esc} = 0.2, f^{\rm{Ly}\alpha}_{\rm esc} = 0.2, E_{\rm UV} = 1.0$ in their equation 8 and 15, and derive the conversion factor:
\begin{equation}
\frac{SFR}{M_\odot/ \rm{yr}} = 2.29 \times 10^{-41} \frac{L_{\rm{Ly}\alpha}}{\rm{erg / s}}.
\end{equation}
For other spectral lines, we adopt the relation from \citet{1998ARA&A..36..189K} and \citet{2007ApJ...657..738L}:
\begin{align}
\frac{SFR}{M_\odot/ \rm{yr}} &= (7.9 \pm 2.4) \times 10^{-42} \frac{L_{H\alpha}}{\rm{erg / s}},\\
\frac{SFR}{M_\odot/ \rm{yr}} &= (1.4 \pm 0.4) \times 10^{-41} \frac{L_{[O II]}}{\rm{erg / s}},\\
\frac{SFR}{M_\odot/ \rm{yr}} &= (7.6 \pm 3.7) \times 10^{-42} \frac{L_{[O III]}}{\rm{erg / s}},
\end{align}
and for the H$\beta$ line, we use the fixed line ratio $H\beta/H\alpha=0.35$ \citep{2006agna.book.....O}.

\bibliography{reference}{}

\begin{thebibliography}{}
\expandafter\ifx\csname natexlab\endcsname\relax\def\natexlab#1{#1}\fi
\providecommand{\url}[1]{\href{#1}{#1}}
\providecommand{\dodoi}[1]{doi:~\href{http://doi.org/#1}{\nolinkurl{#1}}}
\providecommand{\doeprint}[1]{\href{http://ascl.net/#1}{\nolinkurl{http://ascl.net/#1}}}
\providecommand{\doarXiv}[1]{\href{https://arxiv.org/abs/#1}{\nolinkurl{https://arxiv.org/abs/#1}}}

\bibitem[{{Behroozi} {et~al.}(2013){Behroozi}, {Wechsler}, \&
  {Conroy}}]{2013ApJ...770...57B}
{Behroozi}, P.~S., {Wechsler}, R.~H., \& {Conroy}, C. 2013, \apj, 770, 57,
  \dodoi{10.1088/0004-637X/770/1/57}

\bibitem[{{Bowman} {et~al.}(2009){Bowman}, {Morales}, \&
  {Hewitt}}]{2009ApJ...695..183B}
{Bowman}, J.~D., {Morales}, M.~F., \& {Hewitt}, J.~N. 2009, \apj, 695, 183,
  \dodoi{10.1088/0004-637X/695/1/183}

\bibitem[{{Breysse} {et~al.}(2017){Breysse}, {Kovetz}, {Behroozi}, {Dai}, \&
  {Kamionkowski}}]{2017MNRAS.467.2996B}
{Breysse}, P.~C., {Kovetz}, E.~D., {Behroozi}, P.~S., {Dai}, L., \&
  {Kamionkowski}, M. 2017, \mnras, 467, 2996, \dodoi{10.1093/mnras/stx203}

\bibitem[{{Breysse} {et~al.}(2014){Breysse}, {Kovetz}, \&
  {Kamionkowski}}]{2014MNRAS.443.3506B}
{Breysse}, P.~C., {Kovetz}, E.~D., \& {Kamionkowski}, M. 2014, \mnras, 443,
  3506, \dodoi{10.1093/mnras/stu1312}

\bibitem[{{Breysse} {et~al.}(2015){Breysse}, {Kovetz}, \&
  {Kamionkowski}}]{2015MNRAS.452.3408B}
---. 2015, \mnras, 452, 3408, \dodoi{10.1093/mnras/stv1476}

\bibitem[{{Breysse} {et~al.}(2016){Breysse}, {Kovetz}, \&
  {Kamionkowski}}]{2016MNRAS.457L.127B}
---. 2016, \mnras, 457, L127, \dodoi{10.1093/mnrasl/slw005}

\bibitem[{{Breysse} \& {Rahman}(2017)}]{2017MNRAS.468..741B}
{Breysse}, P.~C., \& {Rahman}, M. 2017, \mnras, 468, 741,
  \dodoi{10.1093/mnras/stx451}

\bibitem[{Candes {et~al.}(2006)Candes, Romberg, \&
  Tao}]{Candes:2006:RUP:2263435.2272020}
Candes, E.~J., Romberg, J., \& Tao, T. 2006, IEEE Trans. Inf. Theor., 52, 489,
  \dodoi{10.1109/TIT.2005.862083}

\bibitem[{{Carilli}(2011)}]{2011ApJ...730L..30C}
{Carilli}, C.~L. 2011, \apjl, 730, L30, \dodoi{10.1088/2041-8205/730/2/L30}

\bibitem[{{Chang} {et~al.}(2015){Chang}, {Gong}, {Santos}, {Silva}, {Aguirre},
  {Dor{\'e}}, \& {Pritchard}}]{2015aska.confE...4C}
{Chang}, T.~C., {Gong}, Y., {Santos}, M., {et~al.} 2015, Advancing Astrophysics
  with the Square Kilometre Array (AASKA14), 4.
\newblock \doarXiv{1501.04654}

\bibitem[{{Chang} {et~al.}(2010){Chang}, {Pen}, {Bandura}, \&
  {Peterson}}]{2010Natur.466..463C}
{Chang}, T.-C., {Pen}, U.-L., {Bandura}, K., \& {Peterson}, J.~B. 2010, \nat,
  466, 463, \dodoi{10.1038/nature09187}

\bibitem[{{Chang} {et~al.}(2008){Chang}, {Pen}, {Peterson}, \&
  {McDonald}}]{2008PhRvL.100i1303C}
{Chang}, T.-C., {Pen}, U.-L., {Peterson}, J.~B., \& {McDonald}, P. 2008,
  Physical Review Letters, 100, 091303, \dodoi{10.1103/PhysRevLett.100.091303}

\bibitem[{{Chapman} {et~al.}(2012){Chapman}, {Abdalla}, {Harker}, {Jeli{\'c}},
  {Labropoulos}, {Zaroubi}, {Brentjens}, {de Bruyn}, \&
  {Koopmans}}]{2012MNRAS.423.2518C}
{Chapman}, E., {Abdalla}, F.~B., {Harker}, G., {et~al.} 2012, \mnras, 423,
  2518, \dodoi{10.1111/j.1365-2966.2012.21065.x}

\bibitem[{{Cheng} {et~al.}(2016){Cheng}, {Chang}, {Bock}, {Bradford}, \&
  {Cooray}}]{2016ApJ...832..165C}
{Cheng}, Y.-T., {Chang}, T.-C., {Bock}, J., {Bradford}, C.~M., \& {Cooray}, A.
  2016, \apj, 832, 165, \dodoi{10.3847/0004-637X/832/2/165}

\bibitem[{{Cheng} {et~al.}(2019){Cheng}, {de Putter}, {Chang}, \&
  {Dor{\'e}}}]{2019ApJ...877...86C}
{Cheng}, Y.-T., {de Putter}, R., {Chang}, T.-C., \& {Dor{\'e}}, O. 2019, \apj,
  877, 86, \dodoi{10.3847/1538-4357/ab1b2b}

\bibitem[{{Chung} {et~al.}(2019){Chung}, {Viero}, {Church}, {Wechsler},
  {Alvarez}, {Bond}, {Breysse}, {Cleary}, {Eriksen}, {Foss}, {Gundersen},
  {Harper}, {Ihle}, {Keating}, {Murray}, {Padmanabhan}, {Stein}, {Wehus}, \&
  {COMAP Collaboration}}]{2019ApJ...872..186C}
{Chung}, D.~T., {Viero}, M.~P., {Church}, S.~E., {et~al.} 2019, \apj, 872, 186,
  \dodoi{10.3847/1538-4357/ab0027}

\bibitem[{{Comaschi} \& {Ferrara}(2016)}]{2016MNRAS.455..725C}
{Comaschi}, P., \& {Ferrara}, A. 2016, \mnras, 455, 725,
  \dodoi{10.1093/mnras/stv2339}

\bibitem[{{Crites} {et~al.}(2014){Crites}, {Bock}, {Bradford}, {Chang},
  {Cooray}, {Duband}, {Gong}, {Hailey-Dunsheath}, {Hunacek}, {Koch}, {Li},
  {O'Brient}, {Prouve}, {Shirokoff}, {Silva}, {Staniszewski}, {Uzgil}, \&
  {Zemcov}}]{2014SPIE.9153E..1WC}
{Crites}, A.~T., {Bock}, J.~J., {Bradford}, C.~M., {et~al.} 2014, in \procspie,
  Vol. 9153, Millimeter, Submillimeter, and Far-Infrared Detectors and
  Instrumentation for Astronomy VII, 91531W, \dodoi{10.1117/12.2057207}

\bibitem[{{Croft} {et~al.}(2018){Croft}, {Miralda-Escud{\'e}}, {Zheng},
  {Blomqvist}, \& {Pieri}}]{2018MNRAS.481.1320C}
{Croft}, R. A.~C., {Miralda-Escud{\'e}}, J., {Zheng}, Z., {Blomqvist}, M., \&
  {Pieri}, M. 2018, \mnras, 481, 1320, \dodoi{10.1093/mnras/sty2302}

\bibitem[{{Croft} {et~al.}(2016){Croft}, {Miralda-Escud{\'e}}, {Zheng},
  {Bolton}, {Dawson}, {Peterson}, {York}, {Eisenstein}, {Brinkmann},
  {Brownstein}, {Cen}, {Delubac}, {Font-Ribera}, {Hamilton}, {Lee}, {Myers},
  {Palanque-Delabrouille}, {P{\^a}ris}, {Petitjean}, {Pieri}, {Ross}, {Rossi},
  {Schlegel}, {Schneider}, {Slosar}, {Vazquez}, {Viel}, {Weinberg}, \&
  {Y{\`e}che}}]{2016MNRAS.457.3541C}
{Croft}, R.~A.~C., {Miralda-Escud{\'e}}, J., {Zheng}, Z., {et~al.} 2016,
  \mnras, 457, 3541, \dodoi{10.1093/mnras/stw204}

\bibitem[{{Daddi} {et~al.}(2015){Daddi}, {Dannerbauer}, {Liu}, {Aravena},
  {Bournaud}, {Walter}, {Riechers}, {Magdis}, {Sargent}, {B{\'e}thermin},
  {Carilli}, {Cibinel}, {Dickinson}, {Elbaz}, {Gao}, {Gobat}, {Hodge}, \&
  {Krips}}]{2015A&A...577A..46D}
{Daddi}, E., {Dannerbauer}, H., {Liu}, D., {et~al.} 2015, \aap, 577, A46,
  \dodoi{10.1051/0004-6361/201425043}

\bibitem[{{de Putter} {et~al.}(2014){de Putter}, {Holder}, {Chang}, \&
  {Dore}}]{2014arXiv1403.3727D}
{de Putter}, R., {Holder}, G.~P., {Chang}, T.-C., \& {Dore}, O. 2014, ArXiv
  e-prints.
\newblock \doarXiv{1403.3727}

\bibitem[{{Decarli} {et~al.}(2019){Decarli}, {Walter},
  {G{\'o}nzalez-L{\'o}pez}, {Aravena}, {Boogaard}, {Carilli}, {Cox}, {Daddi},
  {Popping}, {Riechers}, {Uzgil}, {Weiss}, {Assef}, {Bacon}, {Bauer},
  {Bertoldi}, {Bouwens}, {Contini}, {Cortes}, {da Cunha}, {D{\'\i}az-Santos},
  {Elbaz}, {Inami}, {Hodge}, {Ivison}, {Le F{\`e}vre}, {Magnelli}, {Novak},
  {Oesch}, {Rix}, {Sargent}, {Smail}, {Swinbank}, {Somerville}, {van der Werf},
  {Wagg}, \& {Wisotzki}}]{2019ApJ...882..138D}
{Decarli}, R., {Walter}, F., {G{\'o}nzalez-L{\'o}pez}, J., {et~al.} 2019, \apj,
  882, 138, \dodoi{10.3847/1538-4357/ab30fe}

\bibitem[{Donoho(2006)}]{Donoho:2006:CS:2263438.2272089}
Donoho, D.~L. 2006, IEEE Trans. Inf. Theor., 52, 1289,
  \dodoi{10.1109/TIT.2006.871582}

\bibitem[{{Dor{\'e}} {et~al.}(2014){Dor{\'e}}, {Bock}, {Ashby}, {Capak},
  {Cooray}, {de Putter}, {Eifler}, {Flagey}, {Gong}, {Habib}, {Heitmann},
  {Hirata}, {Jeong}, {Katti}, {Korngut}, {Krause}, {Lee}, {Masters},
  {Mauskopf}, {Melnick}, {Mennesson}, {Nguyen}, {{\"O}berg}, {Pullen},
  {Raccanelli}, {Smith}, {Song}, {Tolls}, {Unwin}, {Venumadhav}, {Viero},
  {Werner}, \& {Zemcov}}]{2014arXiv1412.4872D}
{Dor{\'e}}, O., {Bock}, J., {Ashby}, M., {et~al.} 2014, ArXiv e-prints.
\newblock \doarXiv{1412.4872}

\bibitem[{{Fonseca} {et~al.}(2017){Fonseca}, {Silva}, {Santos}, \&
  {Cooray}}]{2017MNRAS.464.1948F}
{Fonseca}, J., {Silva}, M.~B., {Santos}, M.~G., \& {Cooray}, A. 2017, \mnras,
  464, 1948, \dodoi{10.1093/mnras/stw2470}

\bibitem[{{Furlanetto} {et~al.}(2006){Furlanetto}, {Oh}, \&
  {Briggs}}]{2006PhR...433..181F}
{Furlanetto}, S.~R., {Oh}, S.~P., \& {Briggs}, F.~H. 2006, \physrep, 433, 181,
  \dodoi{10.1016/j.physrep.2006.08.002}

\bibitem[{{Gong} {et~al.}(2020){Gong}, {Chen}, \&
  {Cooray}}]{2020ApJ...894..152G}
{Gong}, Y., {Chen}, X., \& {Cooray}, A. 2020, \apj, 894, 152,
  \dodoi{10.3847/1538-4357/ab87a0}

\bibitem[{{Gong} {et~al.}(2012){Gong}, {Cooray}, {Silva}, {Santos}, {Bock},
  {Bradford}, \& {Zemcov}}]{2012ApJ...745...49G}
{Gong}, Y., {Cooray}, A., {Silva}, M., {et~al.} 2012, \apj, 745, 49,
  \dodoi{10.1088/0004-637X/745/1/49}

\bibitem[{{Gong} {et~al.}(2011){Gong}, {Cooray}, {Silva}, {Santos}, \&
  {Lubin}}]{2011ApJ...728L..46G}
{Gong}, Y., {Cooray}, A., {Silva}, M.~B., {Santos}, M.~G., \& {Lubin}, P. 2011,
  \apjl, 728, L46, \dodoi{10.1088/2041-8205/728/2/L46}

\bibitem[{{Gong} {et~al.}(2014){Gong}, {Silva}, {Cooray}, \&
  {Santos}}]{2014ApJ...785...72G}
{Gong}, Y., {Silva}, M., {Cooray}, A., \& {Santos}, M.~G. 2014, \apj, 785, 72,
  \dodoi{10.1088/0004-637X/785/1/72}

\bibitem[{{Helgason} {et~al.}(2012){Helgason}, {Ricotti}, \&
  {Kashlinsky}}]{2012ApJ...752..113H}
{Helgason}, K., {Ricotti}, M., \& {Kashlinsky}, A. 2012, \apj, 752, 113,
  \dodoi{10.1088/0004-637X/752/2/113}

\bibitem[{{Ihle} {et~al.}(2019){Ihle}, {Chung}, {Stein}, {Alvarez}, {Bond},
  {Breysse}, {Cleary}, {Eriksen}, {Foss}, {Gundersen}, {Harper}, {Murray},
  {Padmanabhan}, {Viero}, {Wehus}, \& {COMAP
  Collaboration}}]{2019ApJ...871...75I}
{Ihle}, H.~T., {Chung}, D., {Stein}, G., {et~al.} 2019, \apj, 871, 75,
  \dodoi{10.3847/1538-4357/aaf4bc}

\bibitem[{{Keating} {et~al.}(2016){Keating}, {Marrone}, {Bower}, {Leitch},
  {Carlstrom}, \& {DeBoer}}]{2016ApJ...830...34K}
{Keating}, G.~K., {Marrone}, D.~P., {Bower}, G.~C., {et~al.} 2016, \apj, 830,
  34, \dodoi{10.3847/0004-637X/830/1/34}

\bibitem[{{Keating} {et~al.}(2015){Keating}, {Bower}, {Marrone}, {DeBoer},
  {Heiles}, {Chang}, {Carlstrom}, {Greer}, {Hawkins}, {Lamb}, {Leitch},
  {Miller}, {Muchovej}, \& {Woody}}]{2015ApJ...814..140K}
{Keating}, G.~K., {Bower}, G.~C., {Marrone}, D.~P., {et~al.} 2015, \apj, 814,
  140, \dodoi{10.1088/0004-637X/814/2/140}

\bibitem[{{Kennicutt}(1998)}]{1998ARA&A..36..189K}
{Kennicutt}, Robert~C., J. 1998, Annual Review of Astronomy and Astrophysics,
  36, 189, \dodoi{10.1146/annurev.astro.36.1.189}

\bibitem[{{Kogut} {et~al.}(2015){Kogut}, {Dwek}, \&
  {Moseley}}]{2015ApJ...806..234K}
{Kogut}, A., {Dwek}, E., \& {Moseley}, S.~H. 2015, \apj, 806, 234,
  \dodoi{10.1088/0004-637X/806/2/234}

\bibitem[{{Lagache} {et~al.}(2018){Lagache}, {Cousin}, \&
  {Chatzikos}}]{2018A&A...609A.130L}
{Lagache}, G., {Cousin}, M., \& {Chatzikos}, M. 2018, \aap, 609, A130,
  \dodoi{10.1051/0004-6361/201732019}

\bibitem[{{Li} {et~al.}(2016){Li}, {Wechsler}, {Devaraj}, \&
  {Church}}]{2016ApJ...817..169L}
{Li}, T.~Y., {Wechsler}, R.~H., {Devaraj}, K., \& {Church}, S.~E. 2016, \apj,
  817, 169, \dodoi{10.3847/0004-637X/817/2/169}

\bibitem[{{Lidz} {et~al.}(2011){Lidz}, {Furlanetto}, {Oh}, {Aguirre}, {Chang},
  {Dor{\'e}}, \& {Pritchard}}]{2011ApJ...741...70L}
{Lidz}, A., {Furlanetto}, S.~R., {Oh}, S.~P., {et~al.} 2011, \apj, 741, 70,
  \dodoi{10.1088/0004-637X/741/2/70}

\bibitem[{{Lidz} \& {Taylor}(2016)}]{2016ApJ...825..143L}
{Lidz}, A., \& {Taylor}, J. 2016, \apj, 825, 143,
  \dodoi{10.3847/0004-637X/825/2/143}

\bibitem[{{Lidz} {et~al.}(2009){Lidz}, {Zahn}, {Furlanetto}, {McQuinn},
  {Hernquist}, \& {Zaldarriaga}}]{2009ApJ...690..252L}
{Lidz}, A., {Zahn}, O., {Furlanetto}, S.~R., {et~al.} 2009, \apj, 690, 252,
  \dodoi{10.1088/0004-637X/690/1/252}

\bibitem[{{Liu} \& {Tegmark}(2012)}]{2012MNRAS.419.3491L}
{Liu}, A., \& {Tegmark}, M. 2012, \mnras, 419, 3491,
  \dodoi{10.1111/j.1365-2966.2011.19989.x}

\bibitem[{{Ly} {et~al.}(2007){Ly}, {Malkan}, {Kashikawa}, {Shimasaku}, {Doi},
  {Nagao}, {Iye}, {Kodama}, {Morokuma}, \& {Motohara}}]{2007ApJ...657..738L}
{Ly}, C., {Malkan}, M.~A., {Kashikawa}, N., {et~al.} 2007, \apj, 657, 738,
  \dodoi{10.1086/510828}

\bibitem[{{Madau} {et~al.}(1997){Madau}, {Meiksin}, \&
  {Rees}}]{1997ApJ...475..429M}
{Madau}, P., {Meiksin}, A., \& {Rees}, M.~J. 1997, \apj, 475, 429

\bibitem[{{Mallat} \& {Zhang}(1993)}]{1993ITSP...41.3397M}
{Mallat}, S.~G., \& {Zhang}, Z. 1993, IEEE Transactions on Signal Processing,
  41, 3397, \dodoi{10.1109/78.258082}

\bibitem[{{Mashian} {et~al.}(2015){Mashian}, {Sternberg}, \&
  {Loeb}}]{2015JCAP...11..028M}
{Mashian}, N., {Sternberg}, A., \& {Loeb}, A. 2015, JCAP, 11, 028,
  \dodoi{10.1088/1475-7516/2015/11/028}

\bibitem[{{Masui} {et~al.}(2013){Masui}, {Switzer}, {Banavar}, {Bandura},
  {Blake}, {Calin}, {Chang}, {Chen}, {Li}, {Liao}, {Natarajan}, {Pen},
  {Peterson}, {Shaw}, \& {Voytek}}]{2013ApJ...763L..20M}
{Masui}, K.~W., {Switzer}, E.~R., {Banavar}, N., {et~al.} 2013, \apjl, 763,
  L20, \dodoi{10.1088/2041-8205/763/1/L20}

\bibitem[{{Morales} {et~al.}(2006){Morales}, {Bowman}, \&
  {Hewitt}}]{2006ApJ...648..767M}
{Morales}, M.~F., {Bowman}, J.~D., \& {Hewitt}, J.~N. 2006, \apj, 648, 767,
  \dodoi{10.1086/506135}

\bibitem[{{Moriwaki} {et~al.}(2020){Moriwaki}, {Filippova}, {Shirasaki}, \&
  {Yoshida}}]{2020MNRAS.496L..54M}
{Moriwaki}, K., {Filippova}, N., {Shirasaki}, M., \& {Yoshida}, N. 2020,
  \mnras, 496, L54, \dodoi{10.1093/mnrasl/slaa088}

\bibitem[{{Moustakas} {et~al.}(2006){Moustakas}, {Kennicutt}, \&
  {Tremonti}}]{2006ApJ...642..775M}
{Moustakas}, J., {Kennicutt}, Robert~C., J., \& {Tremonti}, C.~A. 2006, \apj,
  642, 775, \dodoi{10.1086/500964}

\bibitem[{{Murray} {et~al.}(2013){Murray}, {Power}, \&
  {Robotham}}]{2013A&C.....3...23M}
{Murray}, S.~G., {Power}, C., \& {Robotham}, A.~S.~G. 2013, Astronomy and
  Computing, 3, 23, \dodoi{10.1016/j.ascom.2013.11.001}

\bibitem[{{Osterbrock} \& {Ferland}(2006)}]{2006agna.book.....O}
{Osterbrock}, D.~E., \& {Ferland}, G.~J. 2006, {Astrophysics of gaseous nebulae
  and active galactic nuclei}

\bibitem[{{Parsons} {et~al.}(2012){Parsons}, {Pober}, {Aguirre}, {Carilli},
  {Jacobs}, \& {Moore}}]{2012ApJ...756..165P}
{Parsons}, A.~R., {Pober}, J.~C., {Aguirre}, J.~E., {et~al.} 2012, \apj, 756,
  165, \dodoi{10.1088/0004-637X/756/2/165}

\bibitem[{{Planck Collaboration} {et~al.}(2016){Planck Collaboration}, {Ade},
  {Aghanim}, {Arnaud}, {Ashdown}, {Aumont}, {Baccigalupi}, {Banday},
  {Barreiro}, {Bartlett}, {Bartolo}, {Battaner}, {Battye}, {Benabed},
  {Beno{\^\i}t}, {Benoit-L{\'e}vy}, {Bernard}, {Bersanelli}, {Bielewicz},
  {Bock}, {Bonaldi}, {Bonavera}, {Bond}, {Borrill}, {Bouchet}, {Boulanger},
  {Bucher}, {Burigana}, {Butler}, {Calabrese}, {Cardoso}, {Catalano},
  {Challinor}, {Chamballu}, {Chary}, {Chiang}, {Chluba}, {Christensen},
  {Church}, {Clements}, {Colombi}, {Colombo}, {Combet}, {Coulais}, {Crill},
  {Curto}, {Cuttaia}, {Danese}, {Davies}, {Davis}, {de Bernardis}, {de Rosa},
  {de Zotti}, {Delabrouille}, {D{\'e}sert}, {Di Valentino}, {Dickinson},
  {Diego}, {Dolag}, {Dole}, {Donzelli}, {Dor{\'e}}, {Douspis}, {Ducout},
  {Dunkley}, {Dupac}, {Efstathiou}, {Elsner}, {En{\ss}lin}, {Eriksen},
  {Farhang}, {Fergusson}, {Finelli}, {Forni}, {Frailis}, {Fraisse},
  {Franceschi}, {Frejsel}, {Galeotta}, {Galli}, {Ganga}, {Gauthier}, {Gerbino},
  {Ghosh}, {Giard}, {Giraud-H{\'e}raud}, {Giusarma}, {Gjerl{\o}w},
  {Gonz{\'a}lez-Nuevo}, {G{\'o}rski}, {Gratton}, {Gregorio}, {Gruppuso},
  {Gudmundsson}, {Hamann}, {Hansen}, {Hanson}, {Harrison}, {Helou},
  {Henrot-Versill{\'e}}, {Hern{\'a}ndez-Monteagudo}, {Herranz}, {Hildebrand t},
  {Hivon}, {Hobson}, {Holmes}, {Hornstrup}, {Hovest}, {Huang}, {Huffenberger},
  {Hurier}, {Jaffe}, {Jaffe}, {Jones}, {Juvela}, {Keih{\"a}nen}, {Keskitalo},
  {Kisner}, {Kneissl}, {Knoche}, {Knox}, {Kunz}, {Kurki-Suonio}, {Lagache},
  {L{\"a}hteenm{\"a}ki}, {Lamarre}, {Lasenby}, {Lattanzi}, {Lawrence}, {Leahy},
  {Leonardi}, {Lesgourgues}, {Levrier}, {Lewis}, {Liguori}, {Lilje},
  {Linden-V{\o}rnle}, {L{\'o}pez-Caniego}, {Lubin}, {Mac{\'\i}as-P{\'e}rez},
  {Maggio}, {Maino}, {Mandolesi}, {Mangilli}, {Marchini}, {Maris}, {Martin},
  {Martinelli}, {Mart{\'\i}nez-Gonz{\'a}lez}, {Masi}, {Matarrese}, {McGehee},
  {Meinhold}, {Melchiorri}, {Melin}, {Mendes}, {Mennella}, {Migliaccio},
  {Millea}, {Mitra}, {Miville-Desch{\^e}nes}, {Moneti}, {Montier}, {Morgante},
  {Mortlock}, {Moss}, {Munshi}, {Murphy}, {Naselsky}, {Nati}, {Natoli},
  {Netterfield}, {N{\o}rgaard-Nielsen}, {Noviello}, {Novikov}, {Novikov},
  {Oxborrow}, {Paci}, {Pagano}, {Pajot}, {Paladini}, {Paoletti}, {Partridge},
  {Pasian}, {Patanchon}, {Pearson}, {Perdereau}, {Perotto}, {Perrotta},
  {Pettorino}, {Piacentini}, {Piat}, {Pierpaoli}, {Pietrobon}, {Plaszczynski},
  {Pointecouteau}, {Polenta}, {Popa}, {Pratt}, {Pr{\'e}zeau}, {Prunet},
  {Puget}, {Rachen}, {Reach}, {Rebolo}, {Reinecke}, {Remazeilles}, {Renault},
  {Renzi}, {Ristorcelli}, {Rocha}, {Rosset}, {Rossetti}, {Roudier},
  {Rouill{\'e} d'Orfeuil}, {Rowan-Robinson}, {Rubi{\~n}o-Mart{\'\i}n},
  {Rusholme}, {Said}, {Salvatelli}, {Salvati}, {Sandri}, {Santos},
  {Savelainen}, {Savini}, {Scott}, {Seiffert}, {Serra}, {Shellard}, {Spencer},
  {Spinelli}, {Stolyarov}, {Stompor}, {Sudiwala}, {Sunyaev}, {Sutton},
  {Suur-Uski}, {Sygnet}, {Tauber}, {Terenzi}, {Toffolatti}, {Tomasi},
  {Tristram}, {Trombetti}, {Tucci}, {Tuovinen}, {T{\"u}rler}, {Umana},
  {Valenziano}, {Valiviita}, {Van Tent}, {Vielva}, {Villa}, {Wade}, {Wandelt},
  {Wehus}, {White}, {White}, {Wilkinson}, {Yvon}, {Zacchei}, \&
  {Zonca}}]{2016A&A...594A..13P}
{Planck Collaboration}, {Ade}, P.~A.~R., {Aghanim}, N., {et~al.} 2016, \aap,
  594, A13, \dodoi{10.1051/0004-6361/201525830}

\bibitem[{{Popping} {et~al.}(2016){Popping}, {van Kampen}, {Decarli}, {Spaans},
  {Somerville}, \& {Trager}}]{2016MNRAS.461...93P}
{Popping}, G., {van Kampen}, E., {Decarli}, R., {et~al.} 2016, \mnras, 461, 93,
  \dodoi{10.1093/mnras/stw1323}

\bibitem[{{Pullen} {et~al.}(2013){Pullen}, {Chang}, {Dor{\'e}}, \&
  {Lidz}}]{2013ApJ...768...15P}
{Pullen}, A.~R., {Chang}, T.-C., {Dor{\'e}}, O., \& {Lidz}, A. 2013, \apj, 768,
  15, \dodoi{10.1088/0004-637X/768/1/15}

\bibitem[{{Pullen} {et~al.}(2014){Pullen}, {Dor{\'e}}, \&
  {Bock}}]{2014ApJ...786..111P}
{Pullen}, A.~R., {Dor{\'e}}, O., \& {Bock}, J. 2014, \apj, 786, 111,
  \dodoi{10.1088/0004-637X/786/2/111}

\bibitem[{{Riechers} {et~al.}(2019){Riechers}, {Pavesi}, {Sharon}, {Hodge},
  {Decarli}, {Walter}, {Carilli}, {Aravena}, {da Cunha}, {Daddi}, {Dickinson},
  {Smail}, {Capak}, {Ivison}, {Sargent}, {Scoville}, \&
  {Wagg}}]{2019ApJ...872....7R}
{Riechers}, D.~A., {Pavesi}, R., {Sharon}, C.~E., {et~al.} 2019, \apj, 872, 7,
  \dodoi{10.3847/1538-4357/aafc27}

\bibitem[{{Righi} {et~al.}(2008){Righi}, {Hern{\'a}ndez-Monteagudo}, \&
  {Sunyaev}}]{2008A&A...489..489R}
{Righi}, M., {Hern{\'a}ndez-Monteagudo}, C., \& {Sunyaev}, R.~A. 2008, \aap,
  489, 489, \dodoi{10.1051/0004-6361:200810199}

\bibitem[{{Scott} \& {Rees}(1990)}]{1990MNRAS.247..510S}
{Scott}, D., \& {Rees}, M.~J. 1990, \mnras, 247, 510

\bibitem[{{Sheth} {et~al.}(2001){Sheth}, {Mo}, \&
  {Tormen}}]{2001MNRAS.323....1S}
{Sheth}, R.~K., {Mo}, H.~J., \& {Tormen}, G. 2001, \mnras, 323, 1,
  \dodoi{10.1046/j.1365-8711.2001.04006.x}

\bibitem[{{Silva} {et~al.}(2015){Silva}, {Santos}, {Cooray}, \&
  {Gong}}]{2015ApJ...806..209S}
{Silva}, M., {Santos}, M.~G., {Cooray}, A., \& {Gong}, Y. 2015, \apj, 806, 209,
  \dodoi{10.1088/0004-637X/806/2/209}

\bibitem[{{Silva} {et~al.}(2013){Silva}, {Santos}, {Gong}, {Cooray}, \&
  {Bock}}]{2013ApJ...763..132S}
{Silva}, M.~B., {Santos}, M.~G., {Gong}, Y., {Cooray}, A., \& {Bock}, J. 2013,
  \apj, 763, 132, \dodoi{10.1088/0004-637X/763/2/132}

\bibitem[{{Stompor} {et~al.}(2002){Stompor}, {Balbi}, {Borrill}, {Ferreira},
  {Hanany}, {Jaffe}, {Lee}, {Oh}, {Rabii}, {Richards}, {Smoot}, {Winant}, \&
  {Wu}}]{2002PhRvD..65b2003S}
{Stompor}, R., {Balbi}, A., {Borrill}, J.~D., {et~al.} 2002, \prd, 65, 022003,
  \dodoi{10.1103/PhysRevD.65.022003}

\bibitem[{{Sun} {et~al.}(2018){Sun}, {Moncelsi}, {Viero}, {Silva}, {Bock},
  {Bradford}, {Chang}, {Cheng}, {Cooray}, {Crites}, {Hailey-Dunsheath},
  {Uzgil}, {Hunacek}, \& {Zemcov}}]{2018ApJ...856..107S}
{Sun}, G., {Moncelsi}, L., {Viero}, M.~P., {et~al.} 2018, \apj, 856, 107,
  \dodoi{10.3847/1538-4357/aab3e3}

\bibitem[{{Switzer} {et~al.}(2015){Switzer}, {Chang}, {Masui}, {Pen}, \&
  {Voytek}}]{2015ApJ...815...51S}
{Switzer}, E.~R., {Chang}, T.-C., {Masui}, K.~W., {Pen}, U.-L., \& {Voytek},
  T.~C. 2015, \apj, 815, 51, \dodoi{10.1088/0004-637X/815/1/51}

\bibitem[{{Uzgil} {et~al.}(2014){Uzgil}, {Aguirre}, {Bradford}, \&
  {Lidz}}]{2014ApJ...793..116U}
{Uzgil}, B.~D., {Aguirre}, J.~E., {Bradford}, C.~M., \& {Lidz}, A. 2014, \apj,
  793, 116, \dodoi{10.1088/0004-637X/793/2/116}

\bibitem[{{Visbal} \& {Loeb}(2010)}]{2010JCAP...11..016V}
{Visbal}, E., \& {Loeb}, A. 2010, JCAP, 11, 016,
  \dodoi{10.1088/1475-7516/2010/11/016}

\bibitem[{{Walter} {et~al.}(2014){Walter}, {Decarli}, {Sargent}, {Carilli},
  {Dickinson}, {Riechers}, {Ellis}, {Stark}, {Weiner}, {Aravena}, {Bell},
  {Bertoldi}, {Cox}, {Da Cunha}, {Daddi}, {Downes}, {Lentati}, {Maiolino},
  {Menten}, {Neri}, {Rix}, \& {Weiss}}]{2014ApJ...782...79W}
{Walter}, F., {Decarli}, R., {Sargent}, M., {et~al.} 2014, \apj, 782, 79,
  \dodoi{10.1088/0004-637X/782/2/79}

\bibitem[{{Wyithe} \& {Loeb}(2008)}]{2008MNRAS.383..606W}
{Wyithe}, J.~S.~B., \& {Loeb}, A. 2008, \mnras, 383, 606,
  \dodoi{10.1111/j.1365-2966.2007.12568.x}

\bibitem[{{Yue} {et~al.}(2015){Yue}, {Ferrara}, {Pallottini}, {Gallerani}, \&
  {Vallini}}]{2015MNRAS.450.3829Y}
{Yue}, B., {Ferrara}, A., {Pallottini}, A., {Gallerani}, S., \& {Vallini}, L.
  2015, \mnras, 450, 3829, \dodoi{10.1093/mnras/stv933}

\end{thebibliography}
\bibliographystyle{aasjournal}

\end{document}